\documentclass [12pt]{article}
\usepackage{amsmath,amssymb,cite}

\usepackage{float}
\usepackage[section]{placeins}
\usepackage{tabularx}
\usepackage[english]{babel}
\usepackage[dvips]{graphicx}
\usepackage{indentfirst}
\usepackage{epsfig}
\numberwithin{equation}{section}

\begin{document}

\title{
Phase diagrams of the multitrace quartic matrix models of noncommutative $\Phi^4$}

\author{B. Ydri\footnote{Email:ydri@stp.dias.ie}, K. Ramda,  A. Rouag\\
Department of Physics, Faculty of Sciences, Annaba University,\\
 Annaba, Algeria.
}
\maketitle
\abstract{We report a direct and robust calculation, free from ergodic problems, of the non-uniform-to-uniform (stripe) transition line of noncommutative $\Phi^4_2$ by means of an exact Metropolis algorithm applied to the first non-trivial multitrace correction of this theory on the fuzzy sphere. In fact, we reconstruct the entire phase diagram including the Ising, matrix and stripe boundaries together with the triple point. We also report that the measured critical exponents of the Ising transition line agrees with the Onsager values in two dimensions. The triple point is identified as a termination point of the one-cut-to-two-cut transition line and is located at $(\tilde{b},\tilde{c})=(-1.55,0.4)$ which compares favorably with previous Monte Carlo estimate.}

\section{Introduction}
Noncommutative scalar phi-four theory  is a two-parameter model which enjoys three stable phases: i) disordered (symmetric, one-cut, disk) phase, ii) uniform ordered (Ising, broken, asymmetric one-cut) phase and iii) non-uniform ordered (matrix, stripe, two-cut, annulus) phase. This picture is expected to hold for noncommutative/fuzzy phi-four theory in any dimension, and the three phases are all stable and are expected to meet at a triple point. The non-uniform ordered phase \cite{brazovkii} is a full blown nonperturbative manifestation of the perturbative  UV-IR mixing effect \cite{Minwalla:1999px} which is due to the underlying highly non-local matrix degrees of freedom of the noncommutative scalar field.

The phase structure in four dimensions was discussed using the Hartree-Fock approximation  in \cite{Gubser:2000cd} and studied by means of the Monte Carlo method, employing the fuzzy torus \cite{Ambjorn:2000cs} as regulator, in \cite{Ambjorn:2002nj}. 

In two dimensions the theory is renormalizable \cite{Grosse:2003nw}. The regularized theory on the fuzzy sphere \cite{Hoppe:1982,Madore:1991bw} reads
\begin{eqnarray}
S&=&{\rm Tr}\big(a\Phi[L_a,[L_a,{\Phi}]]+b{\Phi}^2+c{\Phi}^4\big).\label{fundamental0}
\end{eqnarray} 
The Laplacian $\Delta={\cal L}_a{\cal L}_a$ defines the underlying geometry, i.e. the metric, of the fuzzy sphere in the sense of  \cite{Connes:1994yd,Frohlich:1993es}. It is found that the collpased parameters are 
\begin{eqnarray}
\tilde{b}=bN^{-3/2}/a~,~\tilde{c}=cN^{-2}/a^2.
\end{eqnarray} 
The above phase structure was confirmed in two dimensions by means of Monte Carlo simulations on the fuzzy sphere in  \cite{GarciaFlores:2009hf,GarciaFlores:2005xc}. The phase diagram is shown on figures (\ref{phase_diagram}). Both figures were generated using the Metropolis algorithm on the fuzzy sphere. In the first figure coupling of the scalar field $\Phi$ to a $U(1)$ gauge field on the fuzzy sphere is included, and as a consequence, we can employ the $U(N)$ gauge symmetry to reduce the scalar sector to only its eigenvalues. In the second figure an approximate Metropolis algorithm, i.e. it does not satisfy detailed balanced, is used.


The problem of the phase structure of fuzzy scalar phi-four was also studied by means of the Monte Carlo method in \cite{Martin:2004un,Panero:2006bx,Medina:2007nv,Das:2007gm,Ydri:2014rea}. The analytic derivation of the phase diagram of noncommutative phi-four on the fuzzy sphere was attempted in \cite{O'Connor:2007ea,Saemann:2010bw,Polychronakos:2013nca,Tekel:2014bta,Nair:2011ux,Tekel:2013vz,Ydri:2014uaa,Steinacker:2005wj}. The related problem of Monte Carlo simulation of noncommutative phi-four on the fuzzy torus, and the fuzzy disc was considered in \cite{Ambjorn:2002nj}, \cite{Bietenholz:2004xs}, and \cite{Lizzi:2012xy} respectively. For a recent study see \cite{Mejia-Diaz:2014lza}.

In this paper, we are interested in studying by means of the Monte Carlo method the first non-trivial multitrace matrix model, quartic in the scalar field, which approximates noncommutative $\Phi^4$ on the fuzzy sphere. The multitrace approach was initiated in \cite{O'Connor:2007ea,Saemann:2010bw}. See also \cite{Ydri:2014uaa} for a review and an extension of this method to the noncommutative Moyal-Weyl plane. For an earlier approach see \cite{Steinacker:2005wj} and for a similar more non-perturbative approach see \cite{Polychronakos:2013nca,Tekel:2014bta,Nair:2011ux,Tekel:2013vz}. The  multitrace expansion is the analogue of the Hopping parameter expansion on the lattice in the sense that we perform a small kinetic term expansion, i.e. expanding in the parameter $a$ of (\ref{fundamental0}), while treating the potential exactly. This should be contrasted with the small interaction expansion of the usual perturbation theory. This technique is expected to capture the matrix transition between disordered and non-uniform ordered phases with arbitrarily increasing accuracy by including more and more terms in the expansion in $a$. From this we can  then infer and/or estimate the position of the triple point. Capturing the Ising transition, and as a consequence the stripe transition, is more subtle and  is only possible, in our opinion,  if we include odd moments in the effective action and do not impose the symmetry $\Phi\longrightarrow -\Phi$.

The effective action obtained in the multitrace approach is a multitrace matrix model, depending on various moments $m_n=Tr M^n$ of an $N\times N$ matrix $M$, which  to the lowest non-trivial order is of the form

\begin{eqnarray}
V&=&{B}Tr M^2+{C}Tr M^4+D\bigg[ Tr M^2\bigg]^2\nonumber\\
&+&B^{'} (Tr M)^2+C^{'} Tr M Tr M^3+D^{'}(Tr M)^4+A^{'}Tr M^2 (Tr M)^2+....\label{fundamental0m}
\end{eqnarray}
The parameters $B$ and $C$ are shifted values of $b$ and $c$ appearing in (\ref{fundamental0}). The primed parameters depend on $a$. The second line includes terms which depend on the odd moments $m_1$ and $m_3$. By diagonalization we obtain therefore the $N$ eigenvalues of $M$ as our independent set of dynamical degrees of freedom with an effective action of the form   

 \begin{eqnarray}
S_{\rm eff}&=&\sum_{i}(b\lambda_i^2+c\lambda_i^4)-\frac{1}{2}\sum_{i\neq j}\ln(\lambda_i-\lambda_j)^2\nonumber\\
&+&\bigg[\frac{r^2}{8}v_{2,1}\sum_{i\ne j}(\lambda_i-\lambda_j)^2+\frac{r^4}{48}v_{4,1}\sum_{i\ne j}(\lambda_i-\lambda_j)^4-\frac{r^4}{24N^2}v_{2,2}\big[\sum_{i\ne j}(\lambda_i-\lambda_j)^2\big]^2+...\bigg].\nonumber\\
\end{eqnarray}
The coefficients $v_{2,1}$, $v_{4,1}$ and $v_{2,2}$ are given by the following two competing calculations found in \cite{O'Connor:2007ea} (Model I) and  \cite{Ydri:2014uaa} (Model II):
\begin{eqnarray}
&&~v_{2,1}=-1~,~v_{4,1}=\frac{3}{2}~,~v_{2,2}=0~,~{\rm Model}~{\rm I}\nonumber\\
&&~v_{2,1}=+1~,~v_{4,1}=0~,~v_{2,2}=\frac{1}{8}~,~{\rm Model}~{\rm II}.
\end{eqnarray}
The result found  in \cite{Ydri:2014uaa} agrees with the non-perturbative result of \cite{Polychronakos:2013nca} and the corrected result of \cite{Saemann:2010bw}. This can also be confirmed by means of Monte Carlo. The first model in the commutative limit $N\longrightarrow \infty$ is therefore a scalar $\Phi^4$ theory on the sphere modulo multi-integral terms. In here, we will study both models by means of Monte Carlo and show that the first model, though incorrect, sustains the uniform ordered phase.  The second model will sustain the uniform ordered phase only if we add to it higher order multitrace corrections.

Since these models depend only on $N$ independent eigenvalues their Monte Carlo sampling by means of the Metropolis algorithm does not suffer from  any ergodic problem and thus what we get in the simulations is really what should exist in the model non-perturbatively. This should be contrasted with the Monte Carlo simulation (via Metropolis, Hybrid Monte Carlo or other method) of (\ref{fundamental0}) which suffers from severe ergodic problems  which do not allow us easy and transparent access to the stripe transition and the triple point \cite{Martin:2004un,Panero:2006bx,Medina:2007nv,Das:2007gm,Ydri:2014rea}. The model I, which sustains the uniform ordered phase, suffers, however, from critical slowing down, for values of $N$ of the order of $N>60$, and thus the use of the Wolf algorithm \cite{Wolff:1988uh} would have been more appropriate. 

Some of our results in this article include:
\begin{itemize}
\item The phase diagram of model I contains the three phases discussed above. The  critical boundaries are determined and the triple point is located.
\item The uniform ordered phase exists in the model I only with the odd terms included. If we assume the symmetry $M\longrightarrow -M$ then the second line of (\ref{fundamental0m}) becomes identically zero and the uniform ordered phase disappears. This is at least true in the domain studied in this article which includes the triple point of fuzzy $\Phi^4$ on the fuzzy sphere and extends to all its phase diagram probed in \cite{GarciaFlores:2009hf,GarciaFlores:2005xc}.
\item The delicate computation of the critical exponents of the Ising transition is discussed and our estimate of the critical exponents $\nu,\alpha,\gamma,\beta$ agrees very well with the Onsager values \cite{Onsager:1943jn}.
\item  The phase diagram of model II, with or without odd terms, does not contain the uniform ordered phase.
\item The one-cut-to-two-cut transition line does not extend to the origin, i.e. to $\tilde{C}=0$,  in the model II which gives us an estimation of the triple point in this case. 
\item In the model II without the odd terms the termination point can be computed analytically from the requirement that the critical point $\tilde{B}_*$ remains always negative and the obtained result  $(\tilde{B},\tilde{C})=(0,1/12)$ agrees with Monte Carlo. 
\item In the model II with odd terms the termination point is found to be located at  $(\tilde{B},\tilde{C})=(-1.05,0.4)$. This is our measurement of the triple point.
\item In all cases the one-cut-to-two-cut matrix transition line agrees better with the doubletrace matrix theory than with the quartic matrix model. The  doubletrace matrix theory is given by $D\neq0$ while all primed parameters are zero.
\item The model of Grosse-Wulkenhaar is also briefly discussed.
\end{itemize}
This article is organized as follows:
\begin{enumerate}
\item{}Section $2$: The Multitrace Matrix Models.
\item{}Section $3$: Exact Solutions.
\begin{itemize}
\item The Pure Real Quartic Matrix Model.
\item The Doubletrace Quartic Matrix Model.
\end{itemize}
\item{}Section $4$: Algorithm.
\item{}Section $5$: Monte Carlo Results.
\begin{itemize}
\item General Remarks.
\item Monte Carlo Tests of Multitrace Approximations.
\item Phase Diagrams.
\item Gross-Wulkenhaar Model.
\end{itemize}
\item{}Section $5$: Conclusion.
\end{enumerate}
We also include appendices for the benefit of interested readers and to make the presentation as self-contained as possible. 
\section{The Multitrace Matrix Models}
Our primary interest here is the theory of noncommutative $\Phi ^4$ on the fuzzy sphere given by the action
\begin{eqnarray}
S=\frac{4\pi R^2}{N+1} Tr\bigg(\frac{1}{2R^2}{\Phi}\Delta{\Phi}+\frac{1}{2}m^2{\Phi}^2+\frac{\lambda}{4!}{\Phi}^4\bigg).
\end{eqnarray}
The Laplacian is $\Delta=[L_a,[L_a,...]]$. Equivalently with the substitution ${\Phi}={\cal M}/\sqrt{2\pi\theta}$, where ${\cal M}=\sum_{i,j=1}^NM_{ij}|i><j|$, this action  reads 
\begin{eqnarray}
S=Tr\bigg(a{\cal M}\Delta {\cal M}+b{\cal M}^2+c{\cal M}^4\bigg).\label{ac}
\end{eqnarray}
The parameters are\footnote{The noncommutativity parameter on the fuzzy sphere is related to the radius of the sphere by $\theta=2R^2/\sqrt{N^2-1}$.}
\begin{eqnarray}
a=\frac{1}{2R^2}~,~b=\frac{1}{2}m^2~,~c=\frac{\lambda}{4!}\frac{1}{2\pi\theta}.
\end{eqnarray}
In terms of the matrix $M$ the action reads
\begin{eqnarray}
S[M]&=&r^2K[M]+Tr\big[b M^2+c M^4\big].
\end{eqnarray}
The kinetic matrix is given by
\begin{eqnarray}
K[M]&=&Tr\bigg[-\Gamma^+M\Gamma M-\frac{1}{N+1}\Gamma_3M\Gamma_3M+EM^2\bigg].
\end{eqnarray}
The matrices $\Gamma$, $\Gamma_3$ and $E$ are given by

\begin{eqnarray}
(\Gamma_3)_{lm}= l{\delta}_{lm}~,~(\Gamma)_{lm}=\sqrt{(m-1)(1- \frac{m}{N+1})}{\delta}_{lm-1}~,~(E)_{lm}=(l-\frac{1}{2}){\delta}_{lm}.
\end{eqnarray}
The relationship between the parameters $a$ and $r^2$ is given by  
\begin{eqnarray}
r^2=2aN
\end{eqnarray}
We start from the path integral  

\begin{eqnarray}
Z&=&\int d M ~\exp\big(-S[M]\big)\nonumber\\
&=&\int d \Lambda~\Delta^2(\Lambda) ~\exp\bigg(-Tr\big(b{\Lambda}^2+c{\Lambda}^4\big)\bigg)\int dU~\exp\bigg(-r^2K[U\Lambda U^{-1}]\bigg).
\end{eqnarray}
The second line involves the diagonalization of the matrix $M$ (more on this below). The calculation of the integral over $U\in U(N)$ is a very long calculation done in \cite{Ydri:2014uaa,O'Connor:2007ea}. The end result is a multi-trace effective potential given by
\begin{eqnarray}
S_{\rm eff}&=&\sum_{i}(b\lambda_i^2+c\lambda_i^4)-\frac{1}{2}\sum_{i\neq j}\ln(\lambda_i-\lambda_j)^2\nonumber\\
&+&\bigg[\frac{r^2}{8}v_{2,1}\sum_{i\ne j}(\lambda_i-\lambda_j)^2+\frac{r^4}{48}v_{4,1}\sum_{i\ne j}(\lambda_i-\lambda_j)^4-\frac{r^4}{24N^2}v_{2,2}\big[\sum_{i\ne j}(\lambda_i-\lambda_j)^2\big]^2+...\bigg].\label{Seff}\nonumber\\
\end{eqnarray}
The coefficients $v$ will be given below.

We will assume now that the parameters $b$, $c$ and $r^2$ scale as
 \begin{eqnarray}
\tilde{a}=\frac{a}{N^{\delta_{a}}}~,~\tilde{b}=\frac{b}{N^{\delta_{b}}}~,~\tilde{c}=\frac{c}{N^{\delta_{c}}}~,~\tilde{r}^2=\frac{r^2}{N^{\delta_{r}}}.
\end{eqnarray}
Further, we will assume a scaling $\delta_{\lambda}$ of the  eigenvalues $\lambda$, viz
 \begin{eqnarray}
\tilde{\lambda}=\frac{\lambda}{N^{\delta_{\lambda}}}.
\end{eqnarray}
It is easy to convince ourselves that  in order for the above effective potential to come out of order $N^2$ we must  have the following values
 \begin{eqnarray}
\delta_{a}=-1-2\delta_{\lambda}~,~\delta_{b}=1-2\delta_{\lambda}~,~\delta_{c}=1-4\delta_{\lambda}~,~\delta_{r}=-2\delta_{\lambda}.\label{collapsed1}
\end{eqnarray}
From the Monte Carlo results of \cite{GarciaFlores:2009hf,GarciaFlores:2005xc}, we know that the scaling behavior of the parameters $b$ and $c$ appearing in the above action on the fuzzy sphere  is given by
 \begin{eqnarray}
\delta_b=\frac{3}{2}~,~\delta_{c}=2.
\end{eqnarray}
By substitution we obtain the other scalings
 \begin{eqnarray}
\delta_{\lambda}=-\frac{1}{4}~,~\delta_a=-\frac{1}{2}~,~\delta_{r}=\frac{1}{2}.
\end{eqnarray}
The problem (\ref{Seff}) is a generalization of the quartic  Hermitian matrix potential model. Indeed, by dropping odd moments, this effective potential corresponds to the matrix model given by

\begin{eqnarray}
V=V_0+\Delta V_0.\label{mmm}
\end{eqnarray}
The classical potential and the even correction $\Delta V_0$ are given by
\begin{eqnarray}
V_0&=&{b}Tr M^2+{c}Tr M^4.
\end{eqnarray}
\begin{eqnarray}
\Delta V_0&=&F^{'} Tr M^2+E^{'} Tr M^4+D\bigg[ Tr M^2\bigg]^2.
\end{eqnarray}
The coefficients $F^{'}$, $E^{'}$ and $D$ are given by
\begin{eqnarray}
F^{'}=\frac{aN^2v_{2,1}}{2}~,~E^{'}=\frac{a^2N^3v_{4,1}}{6}~,~D=-\frac{2\eta a^2N^2}{3}.
\end{eqnarray}
The strength of the multi-trace term $\eta$ is given by
\begin{eqnarray}
\eta=v_{2,2}-\frac{3}{4}v_{4,1}.
\end{eqnarray}
By including terms which involve the odd moments we get the effective potential
\begin{eqnarray}
V&=&V_0+\Delta V_0+\Delta V.\label{mmmf}
\end{eqnarray}
The extra contribution and its coefficients are given by
\begin{eqnarray}
\Delta V&=&B^{'} (Tr M)^2+C^{'} Tr M Tr M^3+D^{'}(Tr M)^4+A^{'}Tr M^2 (Tr M)^2.
\end{eqnarray}
\begin{eqnarray}
B^{'}=-\frac{aN}{2}v_{2,1}~,~C^{'}=-\frac{2a^2N^2}{3}v_{4,1}~,~D^{'}=-\frac{2a^2}{3}v_{2,2}~,~A^{'}=\frac{4a^2N}{3}v_{2,2}. 
\end{eqnarray}
The coefficients $v_{2,1}$, $v_{4,1}$ and $v_{2,2}$ are given by the following two competing calculations found in \cite{O'Connor:2007ea} (Model I) and  \cite{Ydri:2014uaa} (Model II):
\begin{eqnarray}
&&~v_{2,1}=-1~,~v_{4,1}=\frac{3}{2}~,~v_{2,2}=0~,~{\rm Model}~{\rm I}\nonumber\\
&&~v_{2,1}=+1~,~v_{4,1}=0~,~v_{2,2}=\frac{1}{8}~,~{\rm Model}~{\rm II}.\label{comp}
\end{eqnarray}
The difference in the sign of $v_{2,1}$ is probably a typo on the part of \cite{O'Connor:2007ea} while the discrepancy in the values of $v_{4,1}$ and $v_{2,2}$ is more serious and is discussed in \cite{Ydri:2014uaa}. The result found  in \cite{Ydri:2014uaa} agrees with the result of \cite{Polychronakos:2013nca} given by their equation $(2.39)$.  The work \cite{Saemann:2010bw} contains the correct calculation which agrees with both the results of  \cite{Polychronakos:2013nca} and \cite{Ydri:2014uaa}.

The one-cut-to-two-cut transition line in the model (\ref{mmm}) is given by the exact result  \cite{Ydri:2014uaa}
\begin{eqnarray}
\tilde{b}_{*} =-\frac{\tilde{a}}{2}v_{2,1}-2\sqrt{\tilde{c}+\frac{\tilde{a}^2}{6}v_{4,1}}+\frac{4\eta\tilde{a}^2}{3\sqrt{\tilde{c}+\frac{\tilde{a}^2}{6}v_{4,1}}}.
\end{eqnarray}
As point out in \cite{Ydri:2014uaa} this result is new. For a generalization of this result see \cite{Polychronakos:2013nca}. This critical value $\tilde{b}_{*}$ is negative for  
\begin{eqnarray}
\tilde{c}\geq \frac{\tilde{a}^2}{6}(4\eta-v_{4,1}).
\end{eqnarray}

\section{Exact Solutions}
In this section we will give a brief description of the exact solutions of the real quartic matrix model $BTr M^2+CTr M^4$ and the doubletrace real quartic matrix model $BTr M^2+CTr M^4+D (TrM^2)^2$.
\subsection{The Pure Real Quartic Matrix Model}
The phase structure of the pure real quartic matrix model  is studied for example in \cite{Shimamune:1981qf,Brezin:1977sv,eynard, Kawahara:2007eu}. In here we will summarize some of the salient results.

The basic model is given by
\begin{eqnarray}
V&=&BTr M^2+CTr M^4\nonumber\\
&=&\frac{N}{g}(-Tr M^2+\frac{1}{4} Tr M^4).
\end{eqnarray}
\begin{eqnarray}
B=-\frac{N}{g}~,~C=\frac{N}{4g}.
\end{eqnarray}
There are two phases in this case:
\paragraph{Disordered phase (one-cut) for $g\geq g_c$:}

\begin{eqnarray}
\rho(\lambda)&=&\frac{1}{N\pi}(2C\lambda^2+B+C\delta^2)\sqrt{\delta^2-\lambda^2}\nonumber\\
&=&\frac{1}{g\pi}(\frac{1}{2}\lambda^2-1+r^2)\sqrt{4r^2-\lambda^2}.\label{disorder}
\end{eqnarray}
\begin{eqnarray}
-2r\leq \lambda\leq 2r.
\end{eqnarray}
\begin{eqnarray}
r=\frac{1}{2}\delta.
\end{eqnarray}
\begin{eqnarray}
\delta^2&=&\frac{1}{3C}(-B+\sqrt{B^2+12 NC})\nonumber\\
&=&\frac{1}{3}(1+\sqrt{1+3g}).
\end{eqnarray}
\paragraph{Non-uniform ordered phase (two-cut) for $g\le g_c$:}

\begin{eqnarray}
\rho(\lambda)&=&\frac{2C|\lambda|}{N\pi}\sqrt{(\lambda^2-\delta_1^2)(\delta_2^2-\lambda^2)}\nonumber\\
&=&\frac{|\lambda|}{2g\pi}\sqrt{(\lambda^2-r_{-}^2)(r_{+}^2-\lambda^2)}.\label{nonuniform}
\end{eqnarray}
\begin{eqnarray}
r_{-}\leq |\lambda|\leq r_{+}.
\end{eqnarray}
\begin{eqnarray}
r_{-}=\delta_1~,~r_{+}=\delta_2.
\end{eqnarray}
\begin{eqnarray}
r_{\mp}^2&=&\frac{1}{2C}(-B\mp 2\sqrt{NC})\nonumber\\
&=&2(1\mp \sqrt{g}).
\end{eqnarray}
\paragraph{Critical point:}
A third order transition between the above two phases occurs at the critical point 
\begin{eqnarray}
g_c=1\leftrightarrow B_c^2=4NC \leftrightarrow B_c=-2\sqrt{NC}.\label{cr}
\end{eqnarray}
\paragraph{Specific heat:} The behavior of the specific heat across the matrix transition provides also a powerful result against which we can calibrate our algorithms and Monte Carlo simulations. In terms of $\bar{B}=B/B_c$ the specific heat reads in the two phases of the theory as follows
\begin{eqnarray}
&&\frac{C_v}{N^2}=\frac{1}{4}~,~\bar{B}=B/B_c<-1\nonumber\\
&&\frac{C_v}{N^2}=\frac{1}{4}+\frac{2\bar{B}^4}{27}-\frac{\bar{B}}{27}(2\bar{B}^2-3)\sqrt{\bar{B}^2+3}~,~\bar{B}>-1.
\end{eqnarray}
\paragraph{Uniform ordered phase:}
The real quartic matrix model admits also a solution with $Tr M\neq 0$ corresponding to a possible uniform-ordered (Ising) phase. This $U(N)-$like solution can appear only for negative values of the mass parameter $\mu$, and it is constructed, for example, in \cite{Shimamune:1981qf}. It is, however, thought that this solution can not yield to a stable phase without the addition of the kinetic term to the  real quartic matrix model.

The density of eigenvalues in this case is given by
\begin{eqnarray}
\rho(z)=\frac{1}{\pi N}(2C z^2 +2\sigma C z+B +2C\sigma^2+C\tau^2)\sqrt{((\sigma+\tau)-z)(z-(\sigma-\tau))}.
\end{eqnarray}
This is a one-cut solution centered around $\tau$ in the interval $[\sigma-\tau,\sigma+\tau]$ where $\sigma$ and $\tau$ are given by
\begin{eqnarray}
\sigma^2=\frac{1}{10C}(-3B +2\sqrt{B^2-15NC})~,~\tau^2=\frac{1}{15C}(-2B -2\sqrt{B^2-15NC}).
\end{eqnarray}
This solution makes sense only for 
\begin{eqnarray}
B\leq B_c=-\sqrt{15}\sqrt{NC}.
\end{eqnarray}
\subsection{The Doubletrace Quartic Matrix Model}

The doubletrace real quartic matrix model  is given by the multitrace matrix model (\ref{fundamental0m}) with all odd moments set to zero, viz

\begin{eqnarray}
V&=&BTr M^2+CTr M^4+D (Tr M^2)^2.
\end{eqnarray}
The scaling of the parameters is given by
\begin{eqnarray}
\tilde{B}=BN^{-3/2}~,~\tilde{C}=CN^{-2}~,~\tilde{D}=DN^{-1}.
\end{eqnarray} 
The phase structure of this model is very similar to the phase structure of the pure real quartic matrix model outlined in the previous section. See for example \cite{Ydri:2014uaa}. The two stable phases are still given by the disordered (one-cut) phase and the non-uniform-ordered (two-cut) phase separated by a deformation of the line $\tilde{B}_*=-2\sqrt{\tilde{C}}$ given by 
\begin{eqnarray}
\tilde{B}_*=-2\sqrt{\tilde{C}}-\frac{2\tilde{D}}{\sqrt{\tilde{C}}}.\label{pre0}
\end{eqnarray}
For a generalization of this result see \cite{Polychronakos:2013nca}. 

Another important result for us here is the existence of a termination point in the  model of \cite{Ydri:2014uaa} since the critical line does not extend to zero. Indeed, in order for the critical value $\tilde{B}_*$ to be negative one must have $\tilde{C}$ in the range \label{pre}
\begin{eqnarray}
\tilde{C}\geq \tilde{C}_*=\frac{2\eta\tilde{a}^2}{3}=\frac{\tilde{a}^2}{12}.
\end{eqnarray}
Thus the termination point is located at (for $\tilde{a}=1$)
\begin{eqnarray}
(\tilde{B},\tilde{C})=(0,1/12).\label{pre}
\end{eqnarray}

\section{Algorithm}

We start from the potential and the partition function

\begin{eqnarray}
V
&=&Tr\big(B M^2+C M^4\big)+D\big(Tr M^2\big)^2\nonumber\\
&+&B^{'} \big(Tr M\big)^2+C^{'} Tr M Tr M^3+D^{'}\big(Tr M\big)^4+A^{'}Tr M^2 \big(Tr M\big)^2.
\end{eqnarray}

\begin{eqnarray}
Z=\int d M ~\exp\big(-V\big).
\end{eqnarray}
The relationship between the two sets of parameters $\{a,b,c\}$ and $\{B,C,D\}$ is given by
\begin{eqnarray}
B={b}+\frac{aN^2v_{2,1}}{2}~,~C={c}+\frac{a^2N^3v_{4,1}}{6}~,~D=-\frac{2\eta a^2N^2}{3}.
\end{eqnarray}
The collpased parameters are
\begin{eqnarray}
\tilde{B}=\frac{B}{N^{\frac{3}{2}}}=\tilde{b}+\frac{\tilde{a}v_{2,1}}{2}~,~\tilde{C}=\frac{C}{N^2}=\tilde{c}+\frac{\tilde{a}^2v_{4,1}}{6}~,~D=-\frac{2\eta \tilde{a}^2N}{3}.
\end{eqnarray}
Only two of these three parameters are independent. For consistency of the large $N$ limit, we must choose $\tilde{a}$ to be any fixed number. We then choose for simplicity $\tilde{a}=1$ or equivalently $D=-2\eta N/3$\footnote{The authors of \cite{GarciaFlores:2009hf,GarciaFlores:2005xc} chose instead $a=1$. This should not make any difference to the Monte Carlo simulations.}. The other parameters are
\begin{eqnarray}
B^{'}=-\frac{aN}{2}v_{2,1}~,~C^{'}=-\frac{2a^2N^2}{3}v_{4,1}~,~D^{'}=-\frac{2a^2}{3}v_{2,2}~,~A^{'}=\frac{4a^2N}{3}v_{2,2}. 
\end{eqnarray}
We can now diagonalize the scalar matrix $M$ as
\begin{eqnarray}
M=U\Lambda U^{-1}.
\end{eqnarray}
We compute 
\begin{eqnarray}
\delta M=U\bigg(\delta\Lambda +[U^{-1}\delta U,\Lambda]\bigg)U^{-1}.
\end{eqnarray}
Thus (with $U^{-1}\delta U=i\delta V$ being an element of the Lie algebra of SU(N))
\begin{eqnarray}
Tr (\delta M)^2&=&Tr (\delta\Lambda)^2+Tr[U^{-1}\delta U,\Lambda]^2\nonumber\\
&=&\sum_i(\delta\lambda_i)^2+\sum_{i\neq j}(\lambda_i-\lambda_j)^2\delta V_{ij}\delta V_{ij}^*.
\end{eqnarray}
We count $N^2$ real degrees of freedom as there should be. The measure is therefore given by
\begin{eqnarray}
dM&=&\prod_id\lambda_i\prod_{i\neq j}dV_{ij}dV_{ij}^*\sqrt{{\rm det}({\rm metric})}\nonumber\\
&=&\prod_id\lambda_i\prod_{i\neq j}dV_{ij}dV_{ij}^*\sqrt{\prod_{i\neq j}(\lambda_i-\lambda_j)^2}.
\end{eqnarray}
We write this as
\begin{eqnarray}
  d  M= d\Lambda dU \Delta^2(\Lambda).
\end{eqnarray}
The $dU$ is the usual Haar measure over the group SU(N) which is normalized such that $\int dU=1$, whereas the Jacobian $\Delta^2(\Lambda)$ is precisely the so-called Vandermonde determinant defined by
\begin{eqnarray}
\Delta^2(\Lambda)= \prod_{i>j}(\lambda_i-\lambda_j)^2.
\end{eqnarray}
The partition function becomes 
\begin{eqnarray}
Z=\int d \Lambda~\Delta^2(\Lambda) ~\exp\bigg(-Tr\big(B{\Lambda}^2+C{\Lambda}^4\big)-D\bigg(Tr \Lambda^2\bigg)^2\bigg).
\end{eqnarray}
We are therefore dealing with an effective potential given by
\begin{eqnarray}
V_{\rm eff}=B\sum_{i=1}\lambda_i^2+C\sum_{i=1}\lambda_i^4+D\bigg(\sum_{i=1}\lambda_i^2\bigg)^2-\frac{1}{2}\sum_{i\neq j}\ln (\lambda_i-\lambda_j)^2.
\end{eqnarray}
We will use the Metropolis  algorithm to study this model. Under the change $\lambda_i\longrightarrow \lambda_i+h$ of the eigenvalue $\lambda_i$ the above effective potential changes as $V_{\rm eff}\longrightarrow V_{\rm eff}+\Delta V_{i,h}$ where 
\begin{eqnarray}
\Delta V_{i,h}&=&B\Delta S_2+C\Delta S_4+D (2S_2\Delta S_2+\Delta S_2^2)+\Delta S_{\rm Vand}\nonumber\\
&+& B^{'}\Delta S_2^{'}+C^{'}\Delta S_4^{'}+D^{'}\bigg((\Delta S_2^{'})^2+2S_1^2\Delta S_2^{'}\bigg)+A^{'}\bigg((S_1+h)\Delta S_2+hS_2\bigg).\nonumber\\
\end{eqnarray}
The monomials $S_n$ are defined by $S_n=\sum_i\lambda_i^n$ while the variations $\Delta S_n$ and $\Delta S_{\rm Vand}$ are given by
\begin{eqnarray}
\Delta S_2=h^2+2h\lambda_i.
\end{eqnarray}
\begin{eqnarray}
\Delta S_4=6h^2\lambda_i^2+4h\lambda_i^3+4h^3\lambda_i+h^4.
\end{eqnarray}

\begin{eqnarray}
\Delta S_{\rm Vand}=-2\sum_{j \ne i}\ln|1+\frac{h}{\lambda_i-\lambda_j}|.
\end{eqnarray}
\begin{eqnarray}
\Delta S_2^{'}=h^2+2hS_1.
\end{eqnarray}
\begin{eqnarray}
\Delta S_4^{'}=(S_1+h)(3h\lambda_i^2+3h^2\lambda_i+h^3)+hS_3.
\end{eqnarray}


\section{Monte Carlo Results}
\subsection{General Remarks}

\begin{enumerate}
\item We use the statistics $2^{P}+2^{P}\times 2^{P^{'}}$ with $P=15-20$ and $P^{'}=5$ with $N=10-60$ and with the Jackknife method to estimate the error bars. We  can even go further to $N=100$ and beyond but noticed that critical slowing down became a serious obstacle especially in the measurement of critical exponents.
\item Our first test for the validity of our simulations is to look at the Schwinger-Dyson identity given for the full multitrace model (\ref{mmmf}) by
\begin{eqnarray}
<\big(2b Tr M^2+4c Tr M^4+2V_2+4V_4\big)>=N^2.\label{sd}
\end{eqnarray}
The quartic and quadratic pieces $V_2$ and $V_4$ are such that
\begin{eqnarray}
\Delta V_0+\Delta V=V_2+V_4.
\end{eqnarray}
In other words,
\begin{eqnarray}
V_2=F^{'} Tr M^2+B^{'} (Tr M)^2.
\end{eqnarray}
\begin{eqnarray}
V_4=E^{'} Tr M^4+D\bigg[ Tr M^2\bigg]^2+C^{'} Tr M Tr M^3+D^{'}(Tr M)^4+A^{'}Tr M^2 (Tr M)^2.
\end{eqnarray}
\item The second powerful test is to look at the conventional quartic matrix model with $a=0$, viz $V=V_0$. The eigenvalues distributions in the two stable phases (disorder(one-cut) and non-uniform order (two-cut)) as well as the demarcation of their boundary are well known analytically given by the formulas (\ref{disorder}), (\ref{nonuniform}) and (\ref{cr}).
\item Even the quartic multitrace approximation itself can be verified directly in Monte Carlo in order to resolve the ambiguity in the coefficients $v$ between \cite{Ydri:2014uaa} and \cite{O'Connor:2007ea}. We must have as identity the two equations 
\begin{eqnarray}
<a \int dU Tr [L_a,U\Lambda U^{-1}]^2>_{V_0}=<-V_2(\Lambda)>_{V_0}.
\end{eqnarray}
\begin{eqnarray}
<\frac{1}{2}\bigg(a \int dU Tr [L_a,U\Lambda U^{-1}]^2\bigg)^2>_{V_0}=<-V_4(\Lambda)+\frac{1}{2}V_2^2(\Lambda)>_{V_0}.
\end{eqnarray}
The coefficients $v$ appear in the potentials $V_2$ and $V_4$. The expectation values are computed with respect to the conventional quartic matrix model $V_0=V_0(\Lambda)$. 

This test clearly requires the computation of the kinetic term and its square which means in particular that we need to numerically perform the integral over $U$ in the term $\int dU Tr [L_a,U\Lambda U^{-1}]^2$ which is not obvious how to do in any direct way. Equivalently, we can undo the diagonalization in the terms involving the kinetic term to obtain instead the equations
\begin{eqnarray}
<a Tr [L_a,M]^2>_{V_0}=<-V_2>_{V_0}.\label{id1}
\end{eqnarray}
\begin{eqnarray}
<\frac{1}{2}\bigg(a Tr [L_a,M]^2\bigg)^2>_{V_0}=<-V_4+\frac{1}{2}V_2^2>_{V_0}.\label{id2}
\end{eqnarray}
Now the expectation values in the left hand side must be computed with respect to the conventional quartic matrix model $V_0=V_0(M)$ with the full matrix $M=U\Lambda U^{-1}$ instead of the eigenvalues matrix $\Lambda$. The expectation values in the right hand side can be computed either ways.

In other words, the eigenvalues Metropolis algorithm employed in this article  to compute terms such as $<-V_2>_{V_0}$ and $<-V_4+V_2^2/2>_{V_0}$ can not be used to compute the terms $<a Tr [L_a,M]^2>_{V_0}$ and $<\big(a Tr [L_a,M]^2\big)^2/2>_{V_0}$. We use instead the hybrid Monte Carlo algorithm to compute these terms as well as the terms  $<-V_2>_{V_0}$ and $<-V_4+V_2^2/2>_{V_0}$ in order to verify the above equations. This also should be viewed as a counter check for the hybrid Monte Carlo algorithm\footnote{Or a counter check for the eigenvalues Metropolis algorithm depending on which algorithm is more trustworthy. However, we firmly believe that the eigenvalues Metropolis algorithm used here is more robust on all accounts.}  since we can compare the values of $<-V_2>_{V_0}$ and $<-V_4+V_2^2/2>_{V_0}$ obtained using the hybrid Monte Carlo with those obtained using our eigenvalues Metropolis algorithm. We note, in passing, that the Metropolis algorithm employed for the eigenvalues problem here is far more efficient than the hybrid Monte Carlo applied to the same problem without diagonalization. But this we can obviously tolerate for testing purposes.  
\item The most detailed order parameter at our disposal is the eigenvalues distribution of the field/matrix $M$ which behaves in distinct ways in various phases. This behavior mimics their behavior in the conventional quartic matrix model $V_0$, viz
\begin{itemize}
\item The disorder (one-cut) phase is characterized by a single-cut eigenvalues distribution symmetric around $0$ since in this phase $<M>=0$. 
\item The non-uniform order (two-cut) phase is characterized by an eigenvalues distribution symmetric around $0$ but with two disjoint supports since $<M>=\sqrt{-b/2c}~\gamma$ where $\gamma$ is any $N-$dimensional idempotent, i.e. $\gamma^2=1$. This appears for large values of $\tilde{c}$.
\item The uniform order (asymmetric one-cut) phase is characterized by a single-cut eigenvalues distributions centered around a non-zero value since $<M>=\sqrt{-b/2c}~{\bf 1}$. This appears for small values of $\tilde{c}$. 
\end{itemize}
\item The sepcific heat defined with the respect to the multitrace potential is given by
\begin{eqnarray}
C_v=<V^2>-<V>^2.
\end{eqnarray}
The relation of this powerful and most difficult to measure second moment with the specific heat of noncommutative $\Phi^4$ on the fuzzy sphere is discussed in the appendix. In any case, this specific heat is expected to approach the specific heat of the original noncommutative $\Phi^4$ for large values of $\tilde{c}$, and as a consequence it can be used to locate the boundary between one-cut and two-cut as in the conventional quartic matrix model with $a=0$.
\item The magnetization and susceptibility are defined by
\begin{eqnarray}
m=<|Tr M|>~,~\chi=<|Tr M|^2>-<|Tr M|>^2.
\end{eqnarray}
The magnetic susceptibility will exhibit peaks in the second order phase transitions between disorder (one-cut) and uniform order (Ising) and between non-uniform order (two-cut) and uniform order (Ising).
\item The total power and power in the zero mode are defined by
\begin{eqnarray}
P_T=<\frac{1}{N}Tr M^2>~,~P_0=<\big(\frac{1}{N}Tr M)^2>.
\end{eqnarray}
In the Ising (uniform order) phase we will have in particular the very distinguished signal $P=P_0$.
\end{enumerate}
\subsection{Monte Carlo Tests of Multitrace Approximations}
It is quite obvious that resolving the ambiguity between the calculations of \cite{O'Connor:2007ea} and  \cite{Ydri:2014uaa}, summarized in equation (\ref{comp}), is straightforward in Monte Carlo. We only need to show that the two equations (\ref{id1}) and (\ref{id2}) hold as identities in the correct calculation. However, this requires a different algorithm than the eigenvalues Metropolis algorithm used here. Indeed, to solve this problem we need to Monte Carlo sample, both the eigenvalues and the angles of the matrix $M$  using the Metropolis or the hybrid Monte Carlo, the quartic matrix model
\begin{eqnarray}
V_0&=&{b}Tr M^2+{c}Tr M^4.
\end{eqnarray}
Monte Carlo simulations of this model can also be compared to the exact solution outlined in section $3.1$ so calibration in this case is easy. The detail of this simple exercise is reported in \cite{Ydri:2014uaa}. There, it is decisively shown that the calculation of  \cite{Ydri:2014uaa} gives the correct approximation of noncommutative scalar $\Phi_2^4$ on the fuzzy sphere.
\subsection{Phase Diagrams}

\begin{figure}[htbp]
\begin{center}
\includegraphics[width=5.0cm,angle=-90]{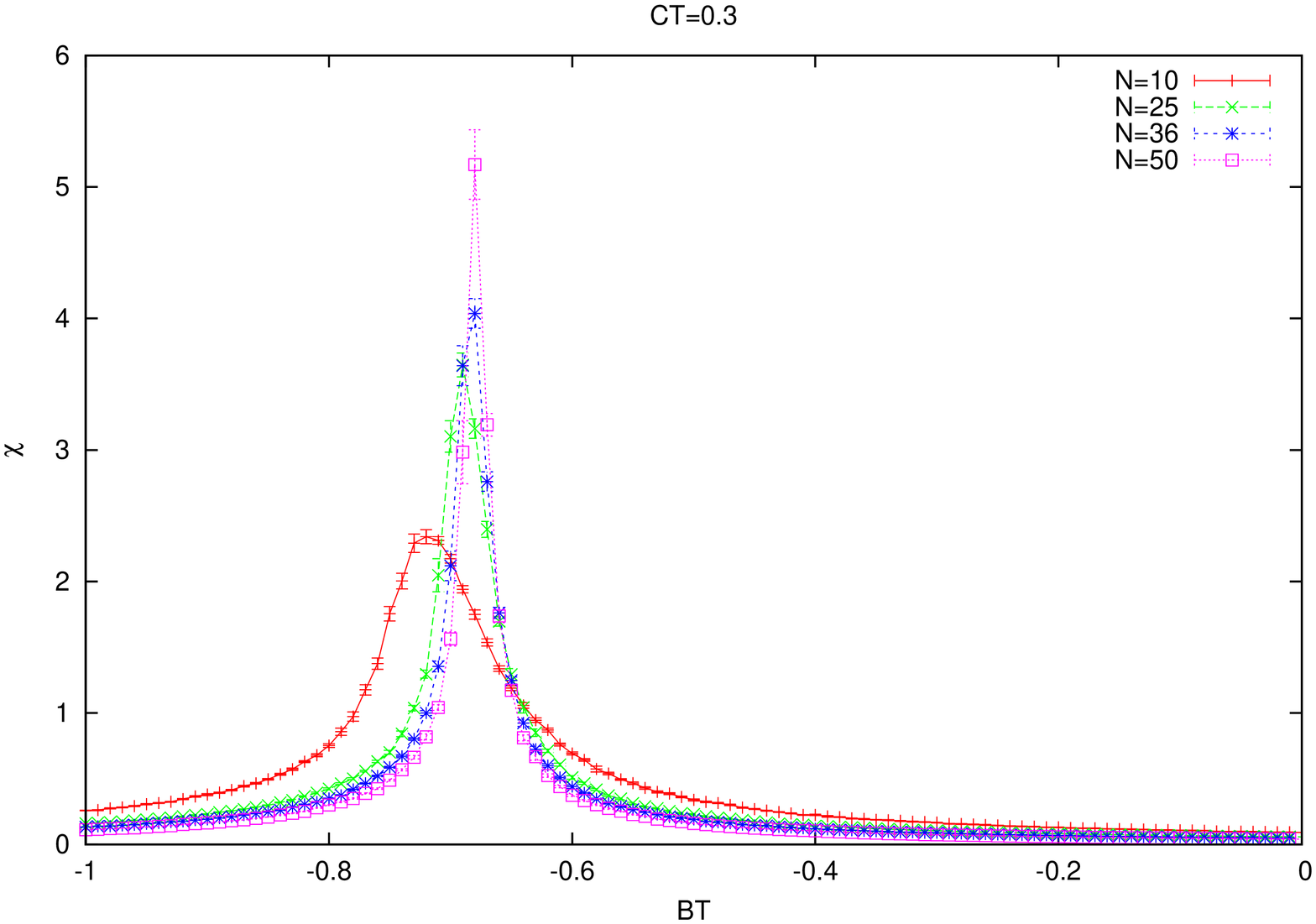}
\includegraphics[width=5.0cm,angle=-90]{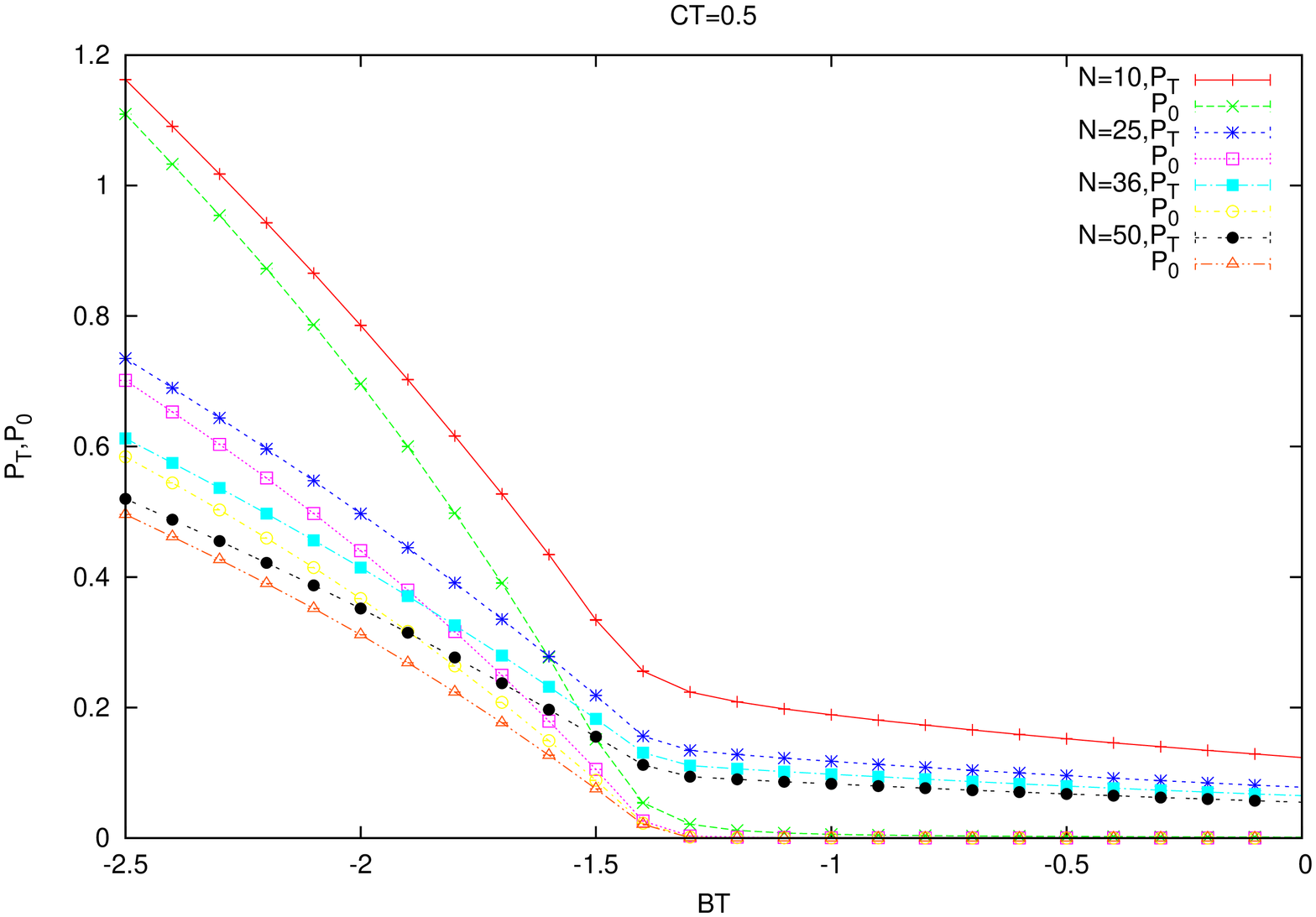}
\includegraphics[width=5.0cm,angle=-90]{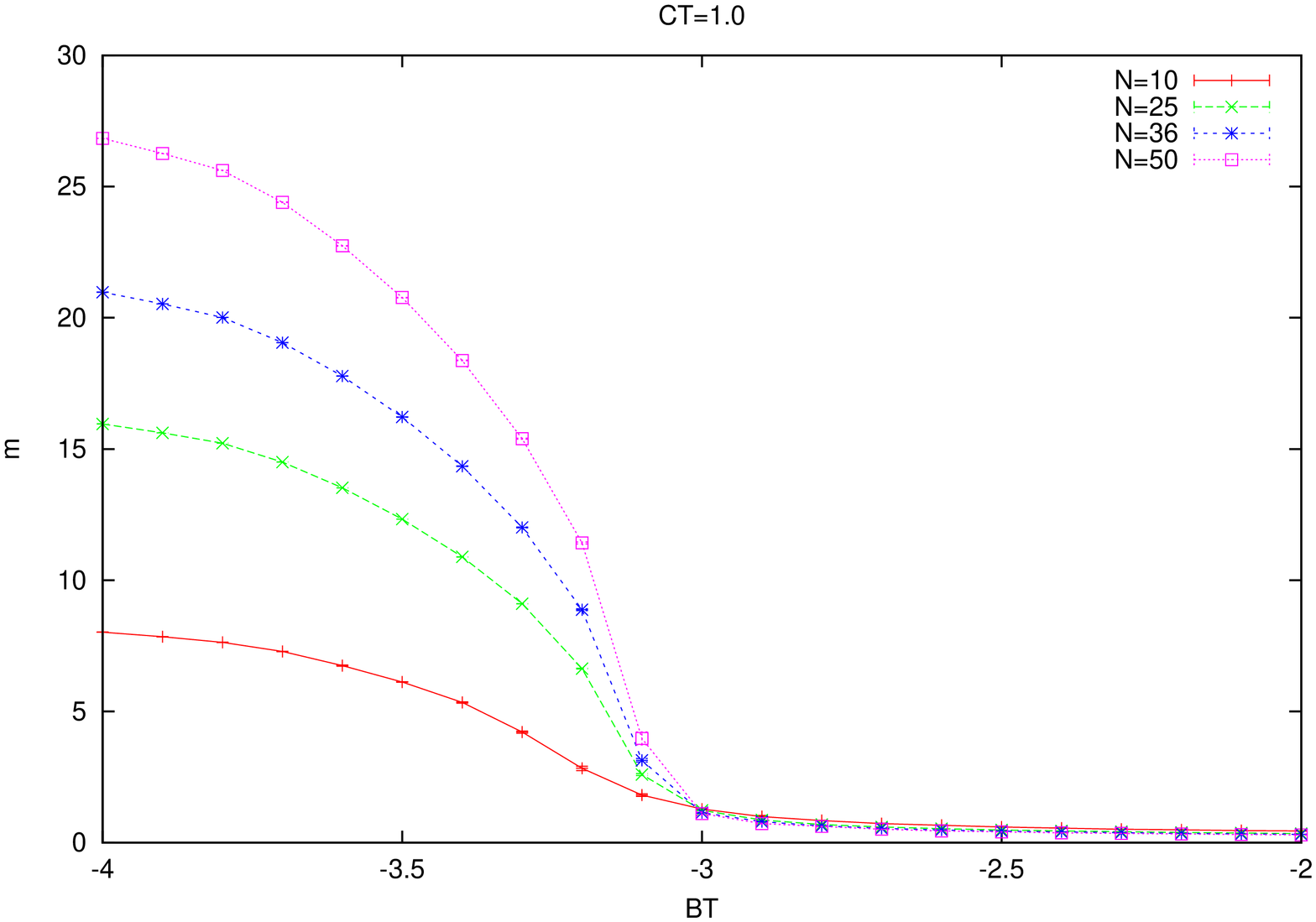}
\includegraphics[width=5.0cm,angle=-90]{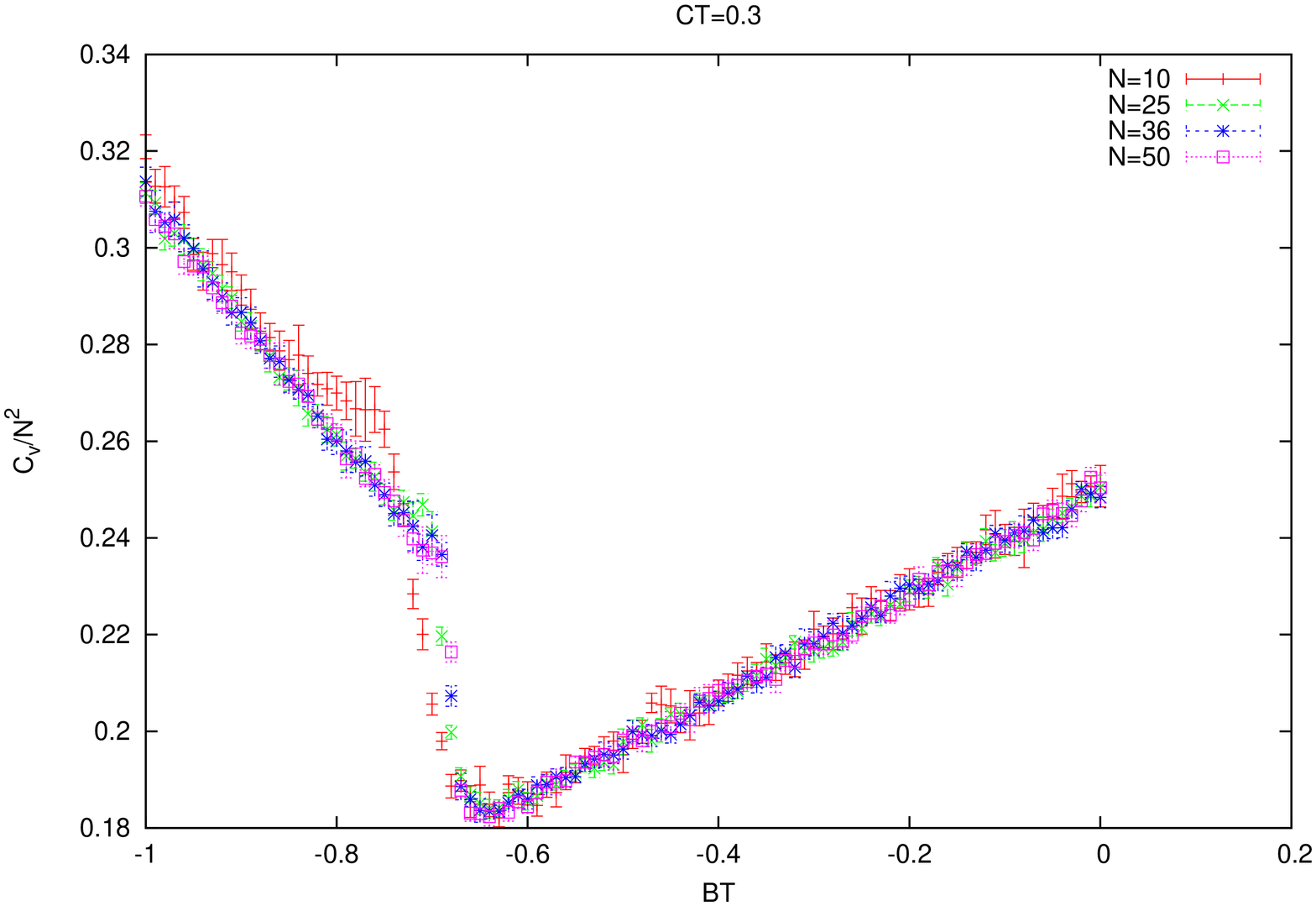}
\end{center}
\caption{Some observables of the multitrace model of  \cite{O'Connor:2007ea} across the disorder-to-uniform transition.}
\label{fig1}
\end{figure}

\begin{figure}[htbp]
\begin{center}
\includegraphics[width=6.0cm,angle=-90]{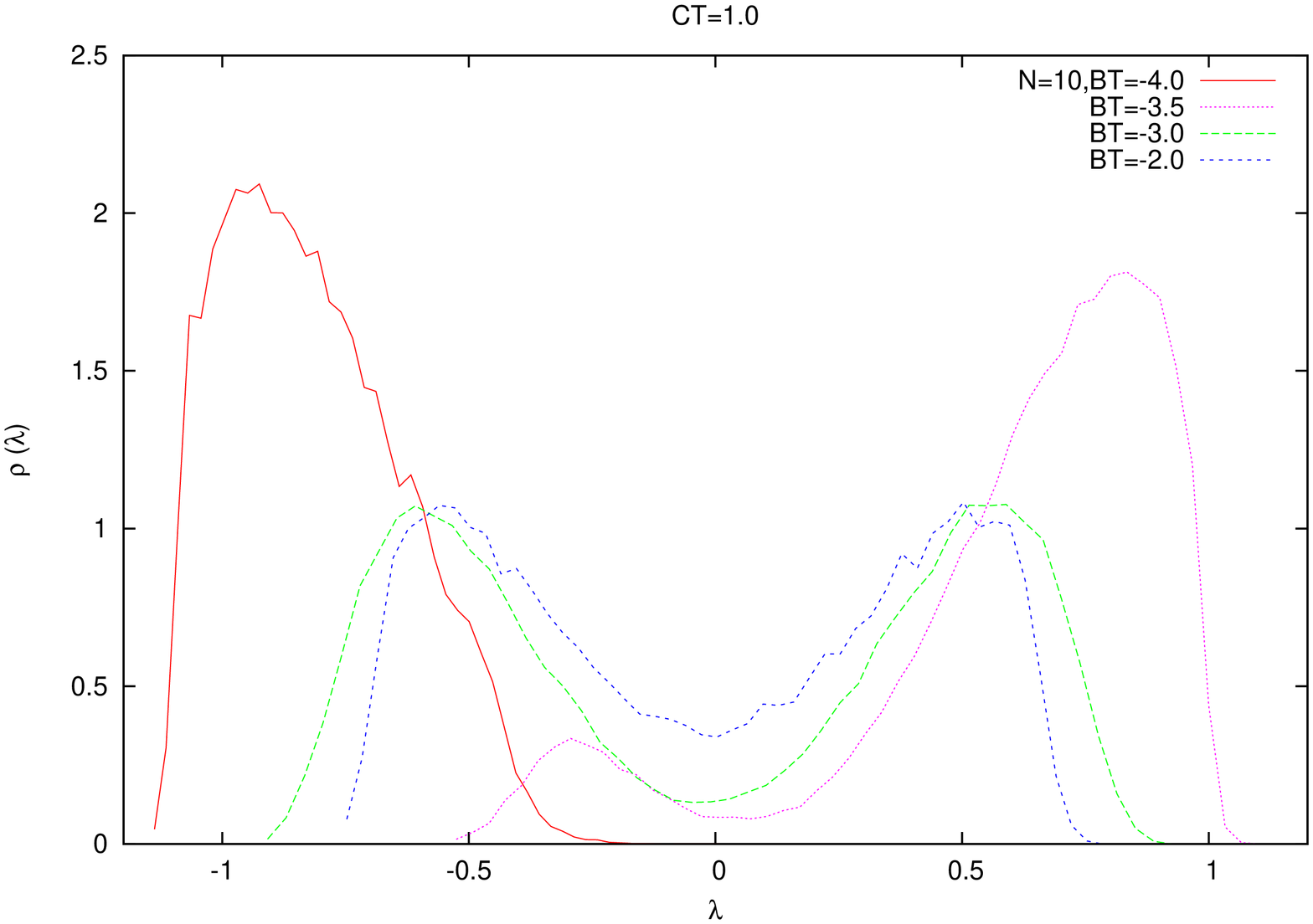}
\includegraphics[width=6.0cm,angle=-90]{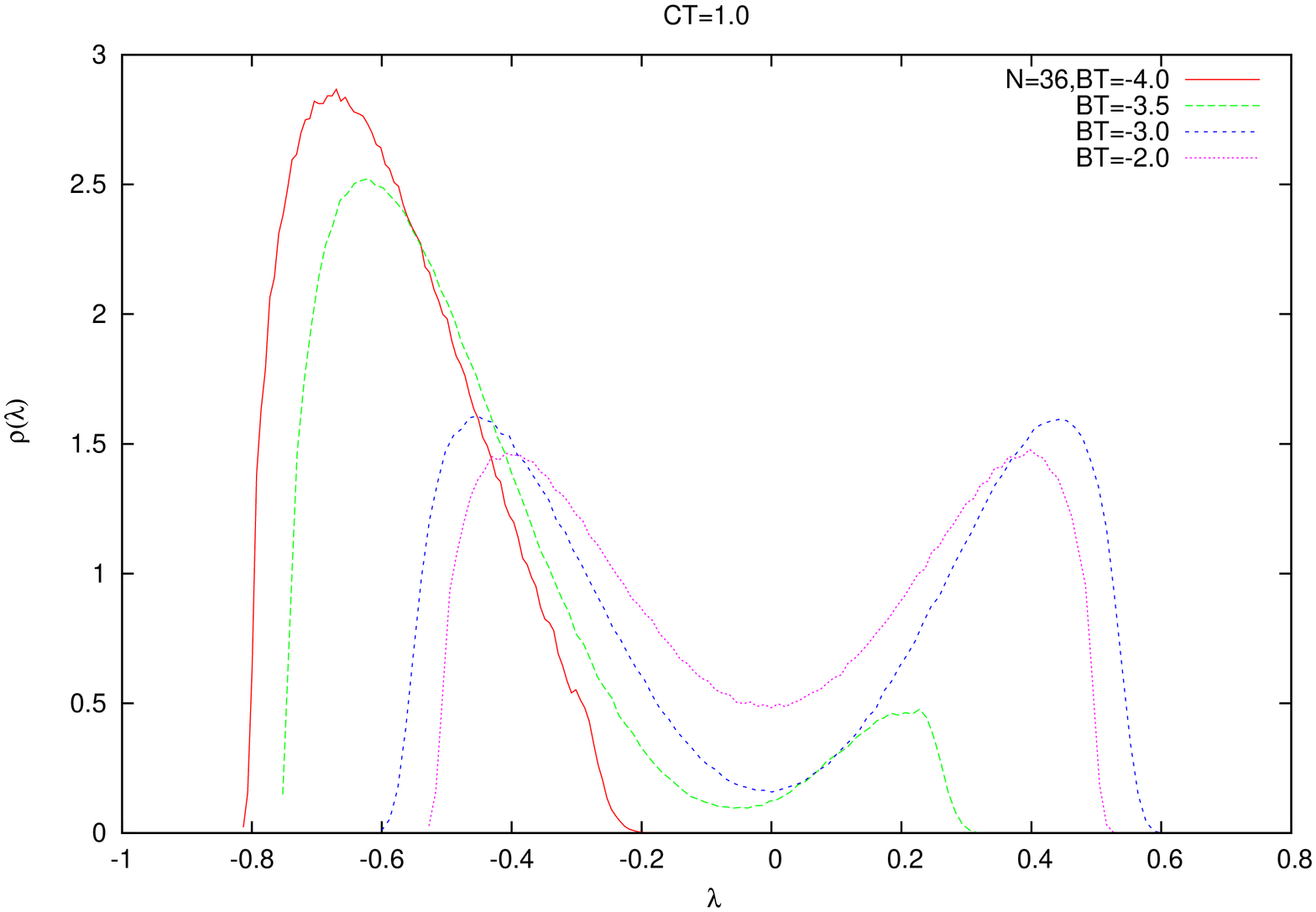}
\end{center}
\caption{The eigenvalues distribution of the matrix $M$ in the multitrace model of  \cite{O'Connor:2007ea} with  $\tilde{C}=1.0$ across the disorder-to-uniform transition.}\label{fig12}
\end{figure}

\begin{figure}[htbp]
\begin{center}
\includegraphics[width=6.0cm,angle=-90]{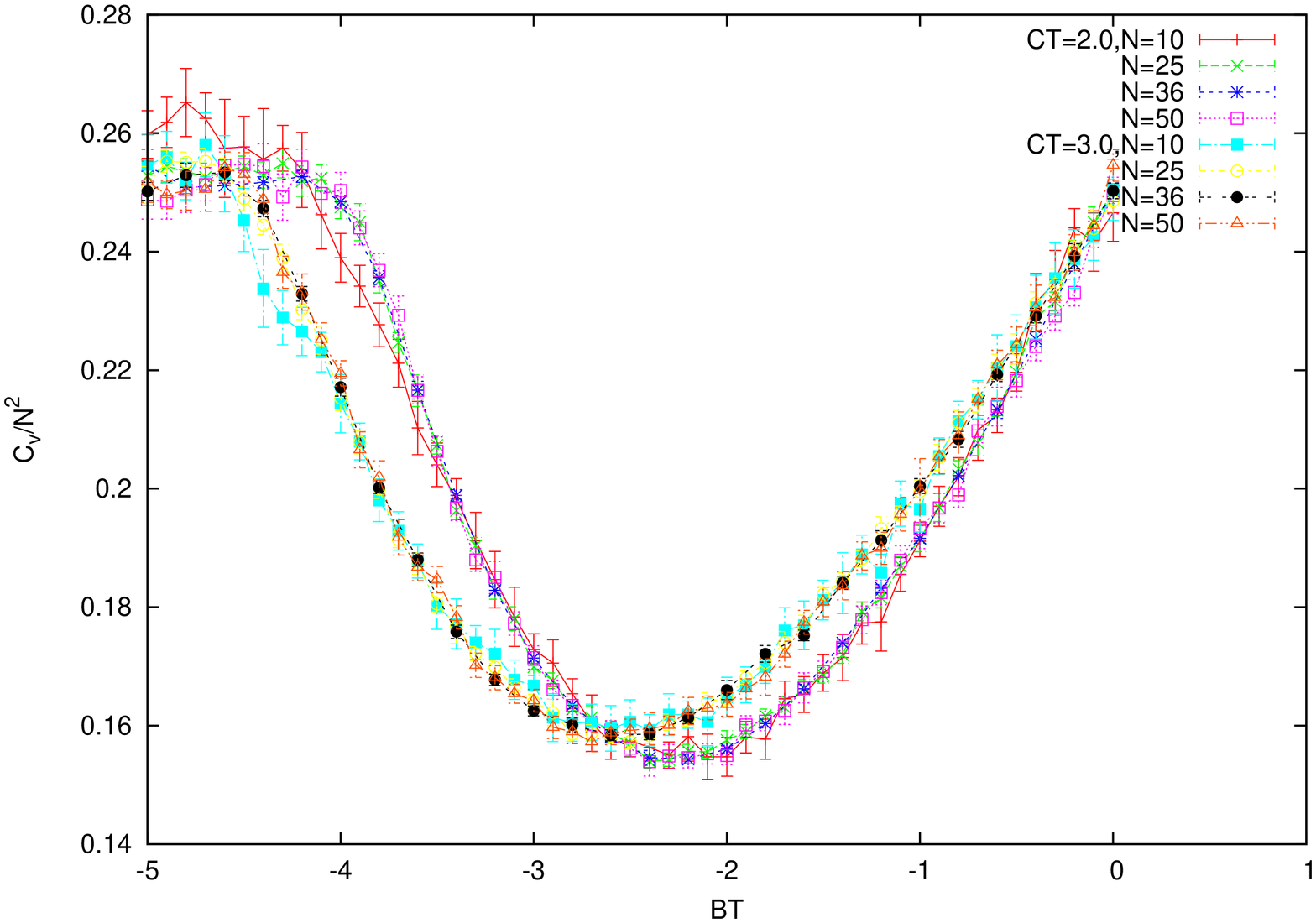}
\includegraphics[width=6.0cm,angle=-90]{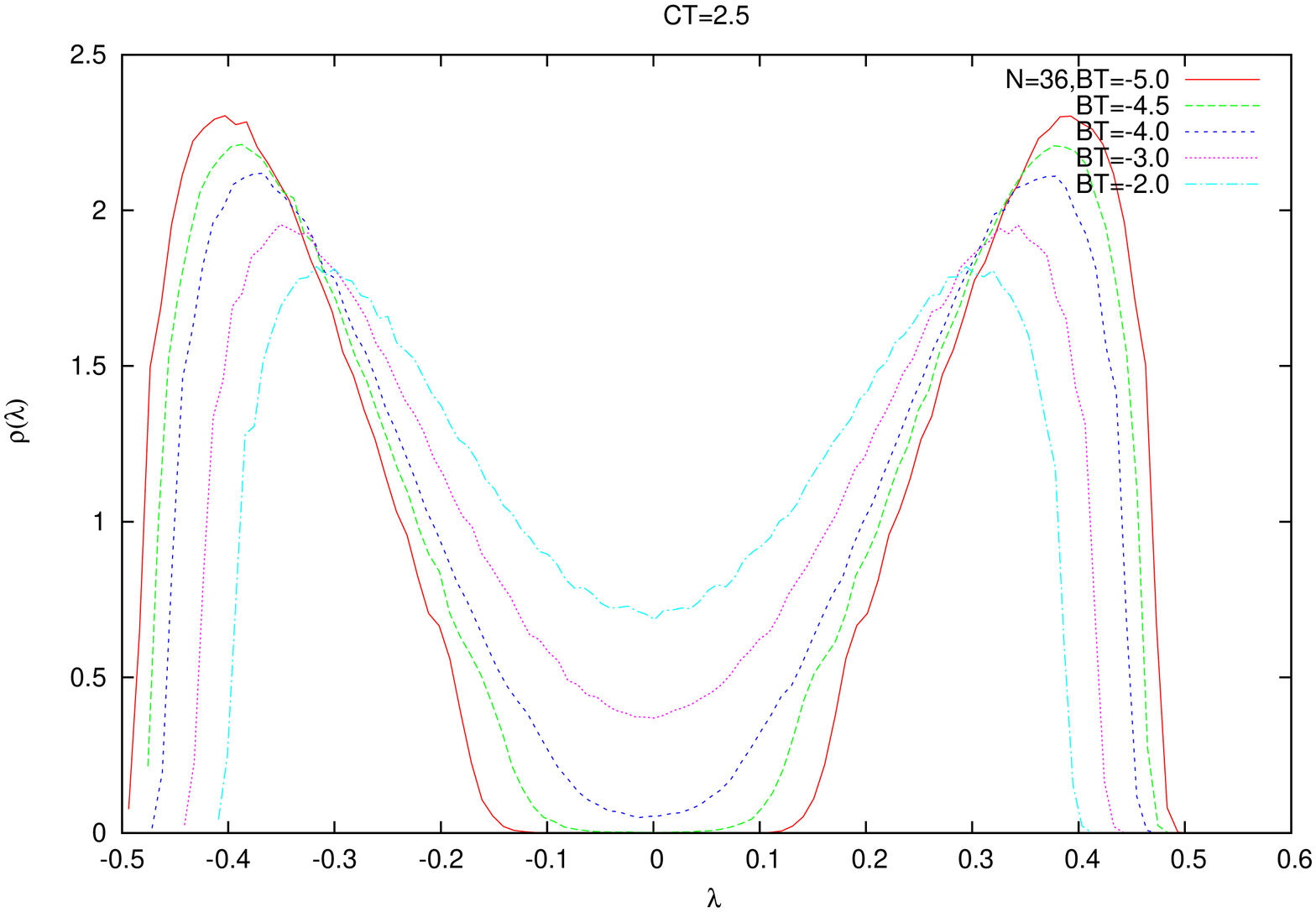}
\end{center}
\caption{The specific heat and the eigenvalues distribution of the matrix $M$ in the multitrace model of  \cite{O'Connor:2007ea} across the disorder-to-non-uniform transition.}\label{fig2}
\end{figure}

\begin{figure}[htbp]
\begin{center}
\includegraphics[width=6.0cm,angle=-90]{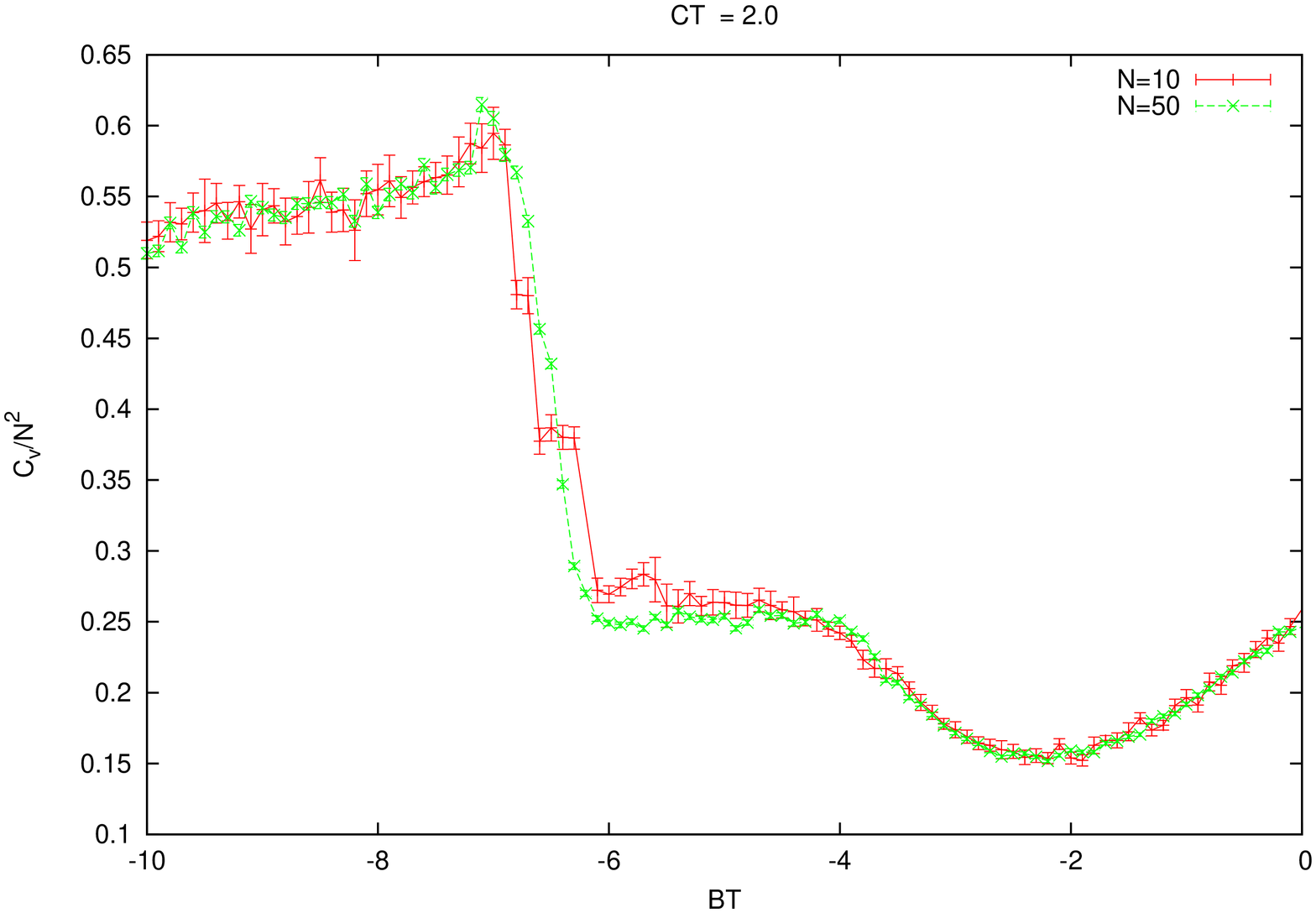}
\includegraphics[width=6.0cm,angle=-90]{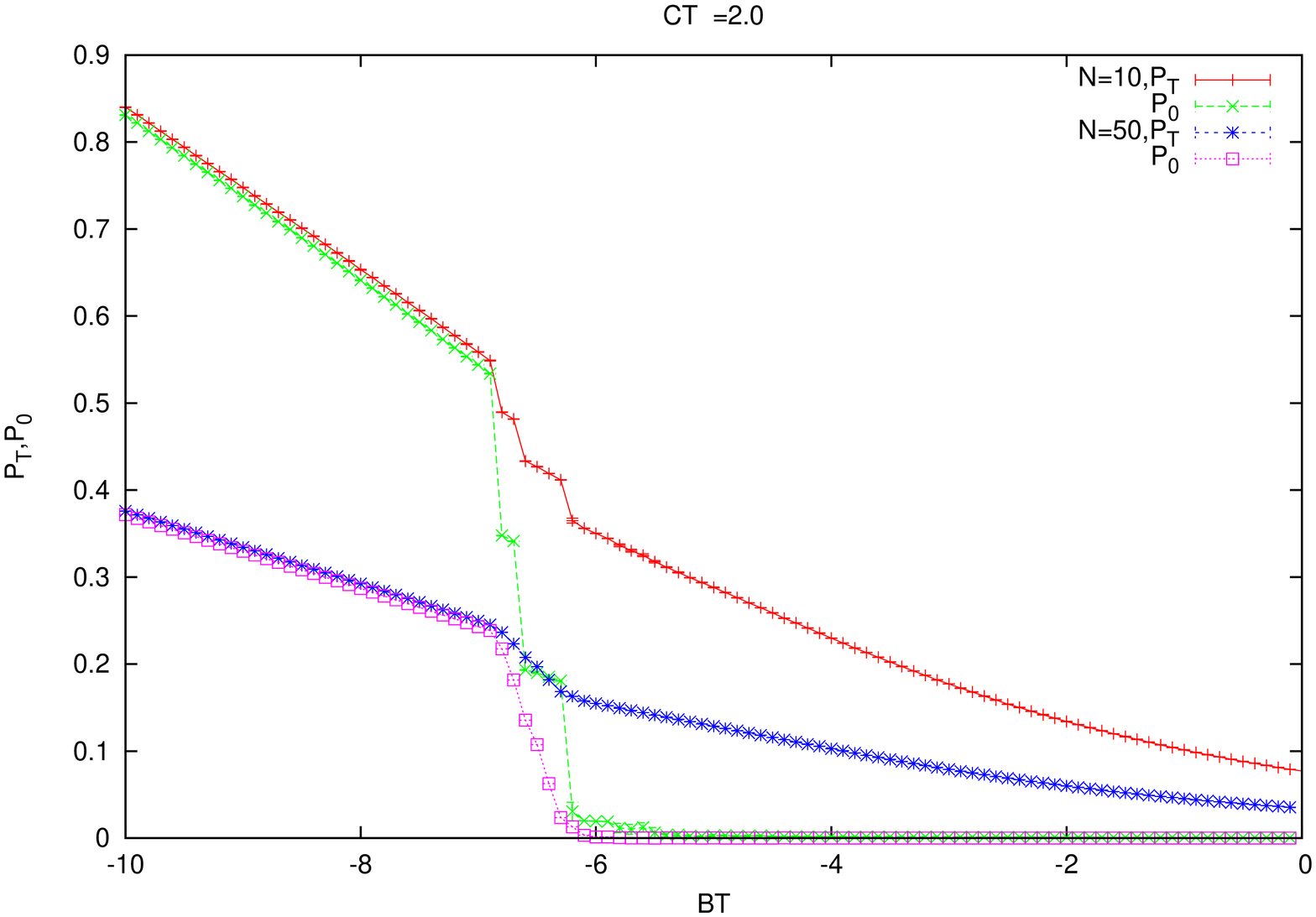}
\end{center}
\caption{The total and zero powers and the specific heat, as functions of $\tilde{B}$, of the multitrace model of  \cite{O'Connor:2007ea} across the non-uniform-to-uniform transition.}
\label{fig3}
\end{figure}

\begin{figure}[htbp]
\begin{center}
\includegraphics[width=6.0cm,angle=-90]{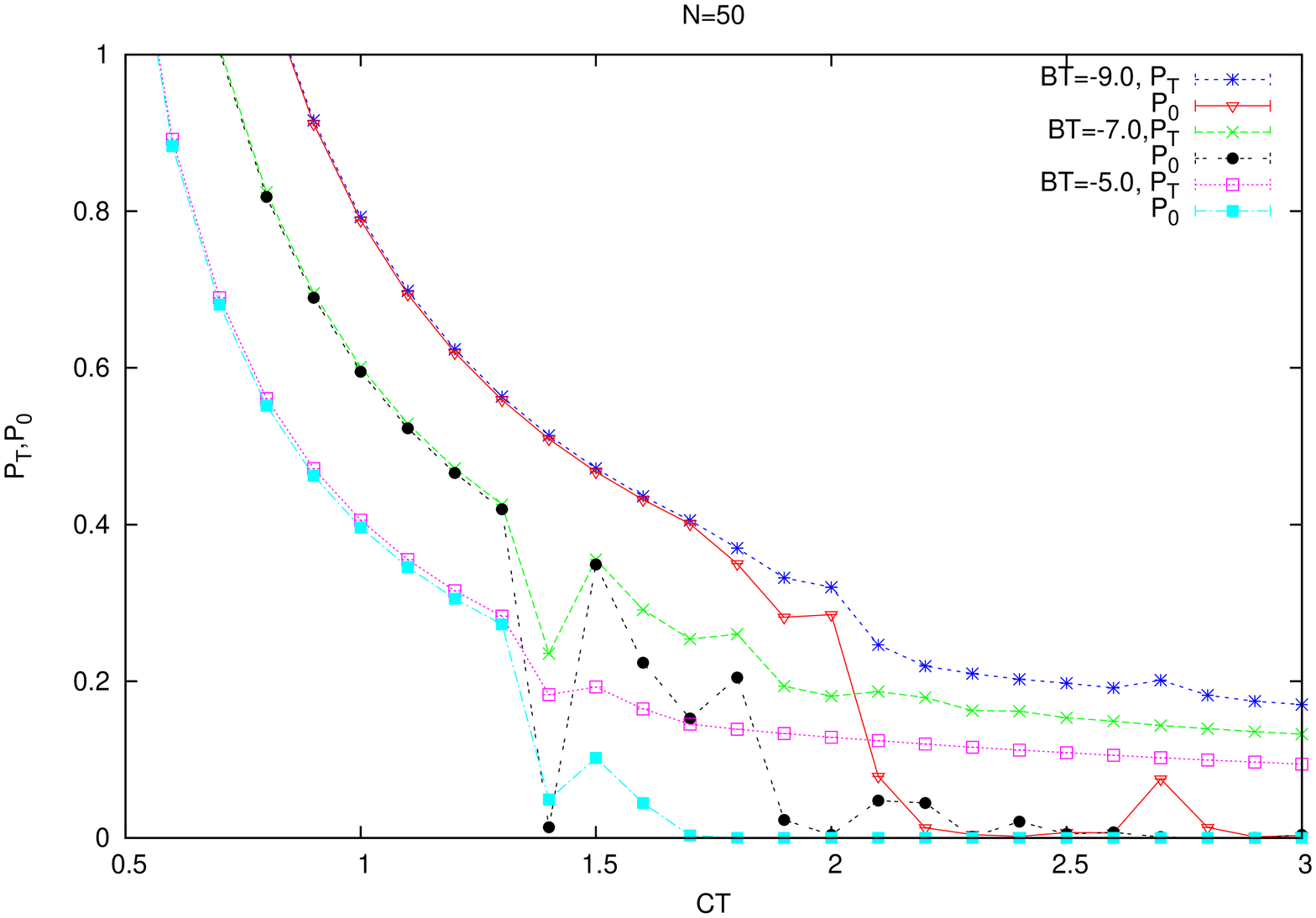}
\includegraphics[width=6.0cm,angle=-90]{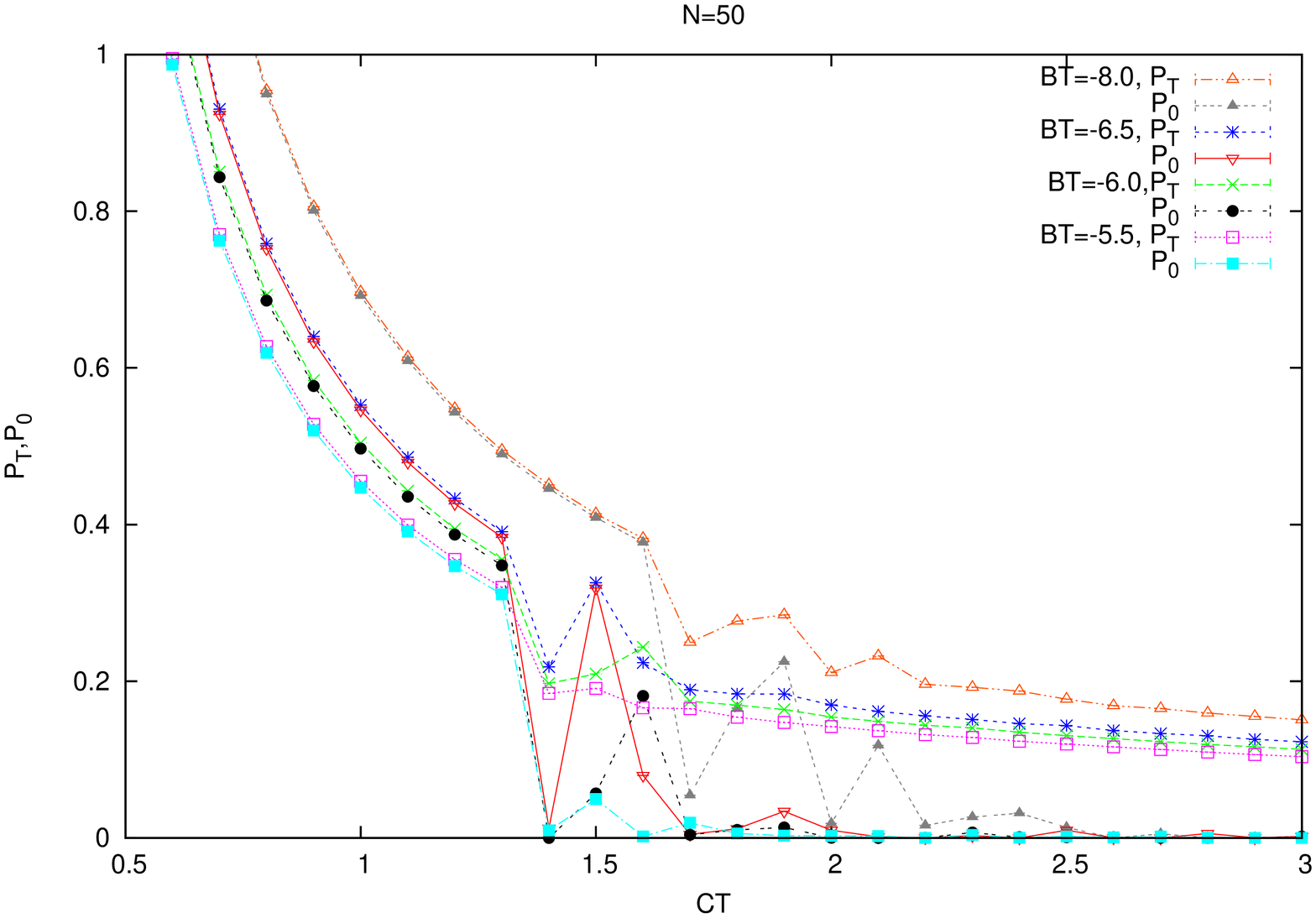}
\includegraphics[width=6.0cm,angle=-90]{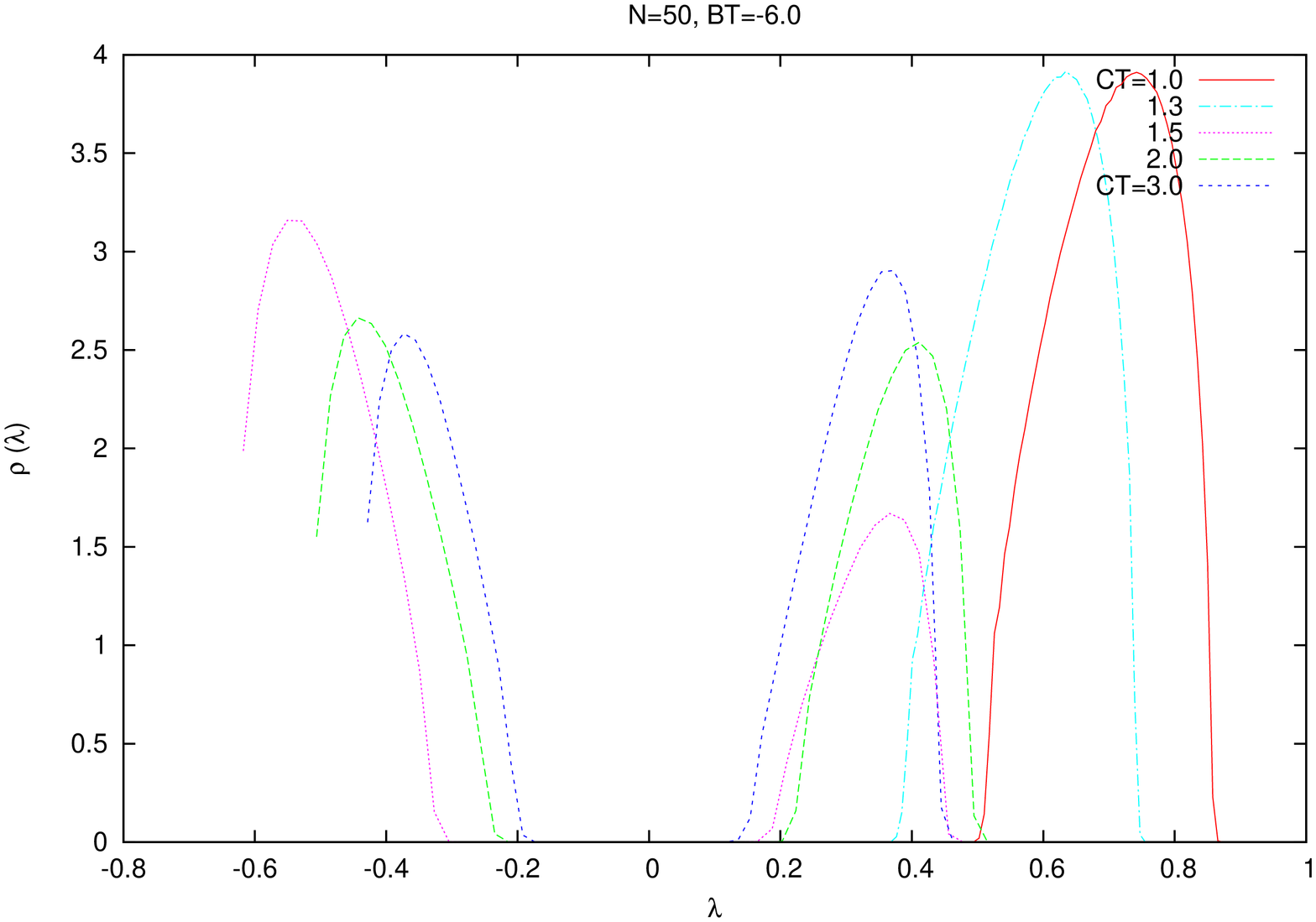}
\end{center}
\caption{The total and zero powers, as functions of $\tilde{C}$, and the eigenvalues distribution of the multitrace model of  \cite{O'Connor:2007ea} across the non-uniform-to-uniform transition.}
\label{fig3e}
\end{figure}

\begin{figure}[htbp]
\begin{center}
\includegraphics[width=9.0cm,angle=-90]{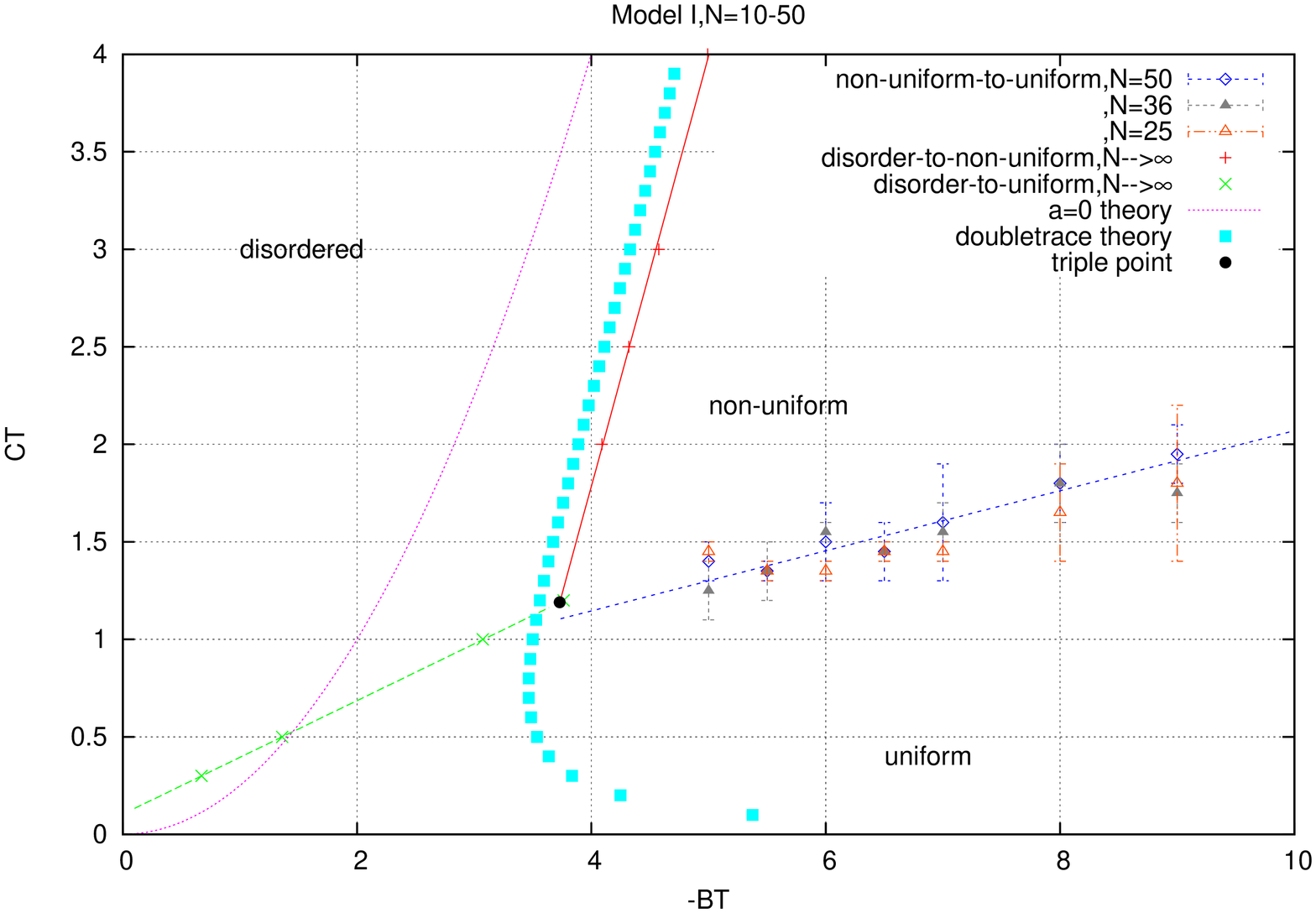}
\includegraphics[width=9.0cm,angle=-90]{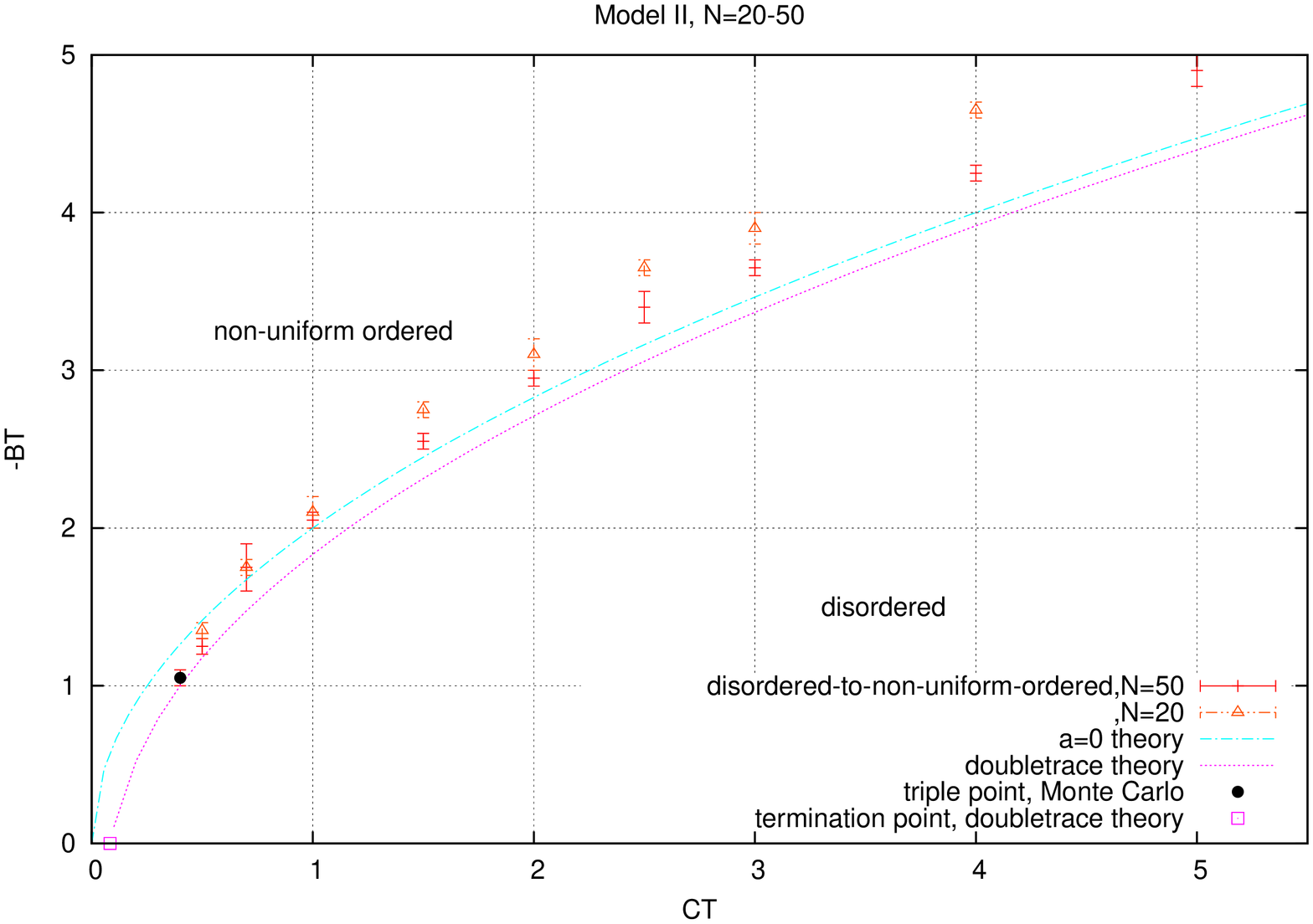}
\end{center}
\caption{The phase diagrams of the multitrace models of  \cite{O'Connor:2007ea} and \cite{Ydri:2014uaa}. Model I: The Ising and matrix transition data points are not shown but only their extrapolated fits are included whereas the $N=25$, $N=36$ and $N=50$ stripe data points are indicated explicitly. Model II: the triple point is identified as the termination point located at $(\tilde{B},\tilde{C})=(-1.05,0.4)$ which is to be compared with the doubletrace prediction at $(0,1/12)$. }\label{pd}
\end{figure}

\begin{figure}[htbp]
\begin{center}
\includegraphics[width=5.0cm,angle=-90]{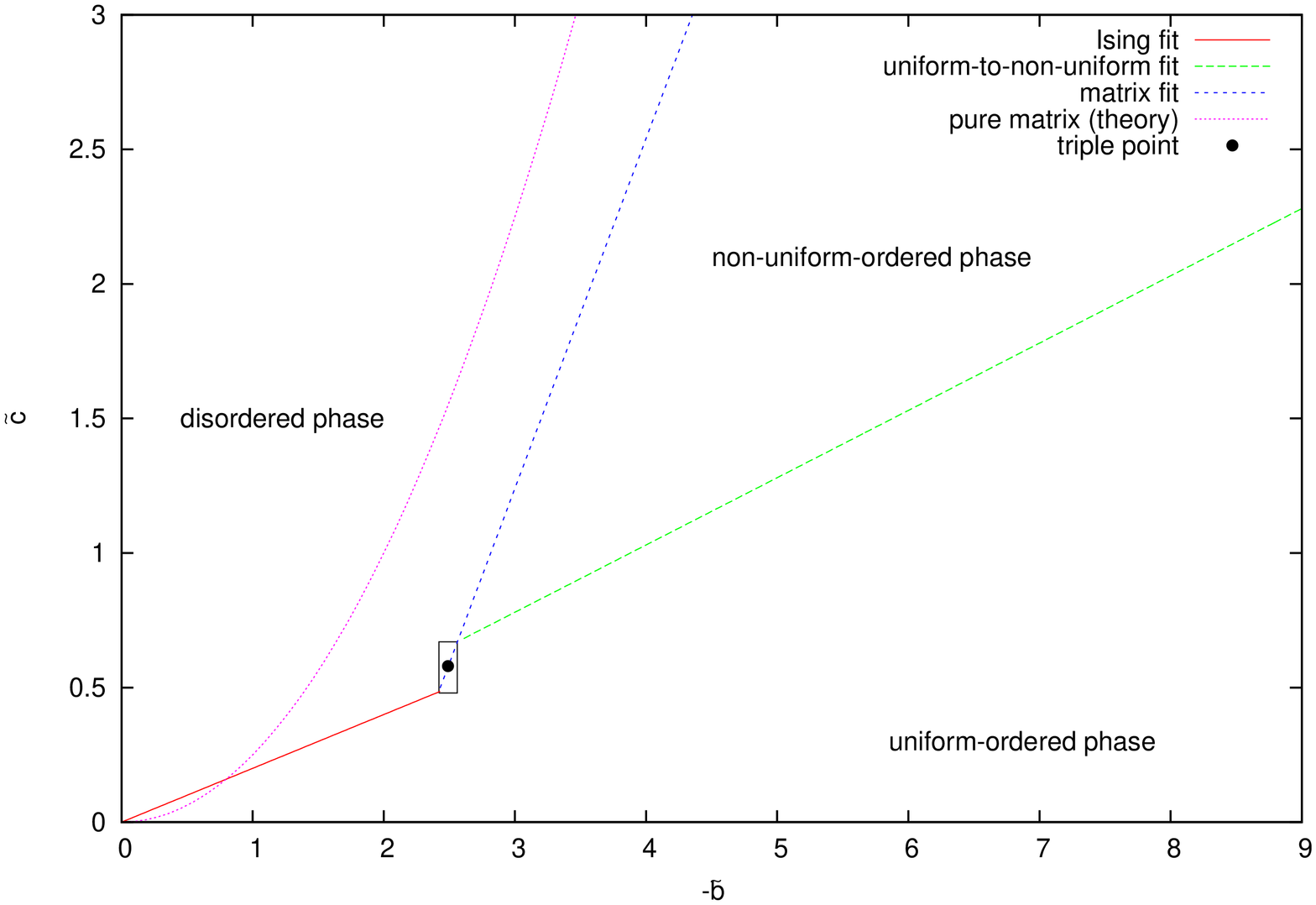}
\includegraphics[width=5.0cm,angle=-90]{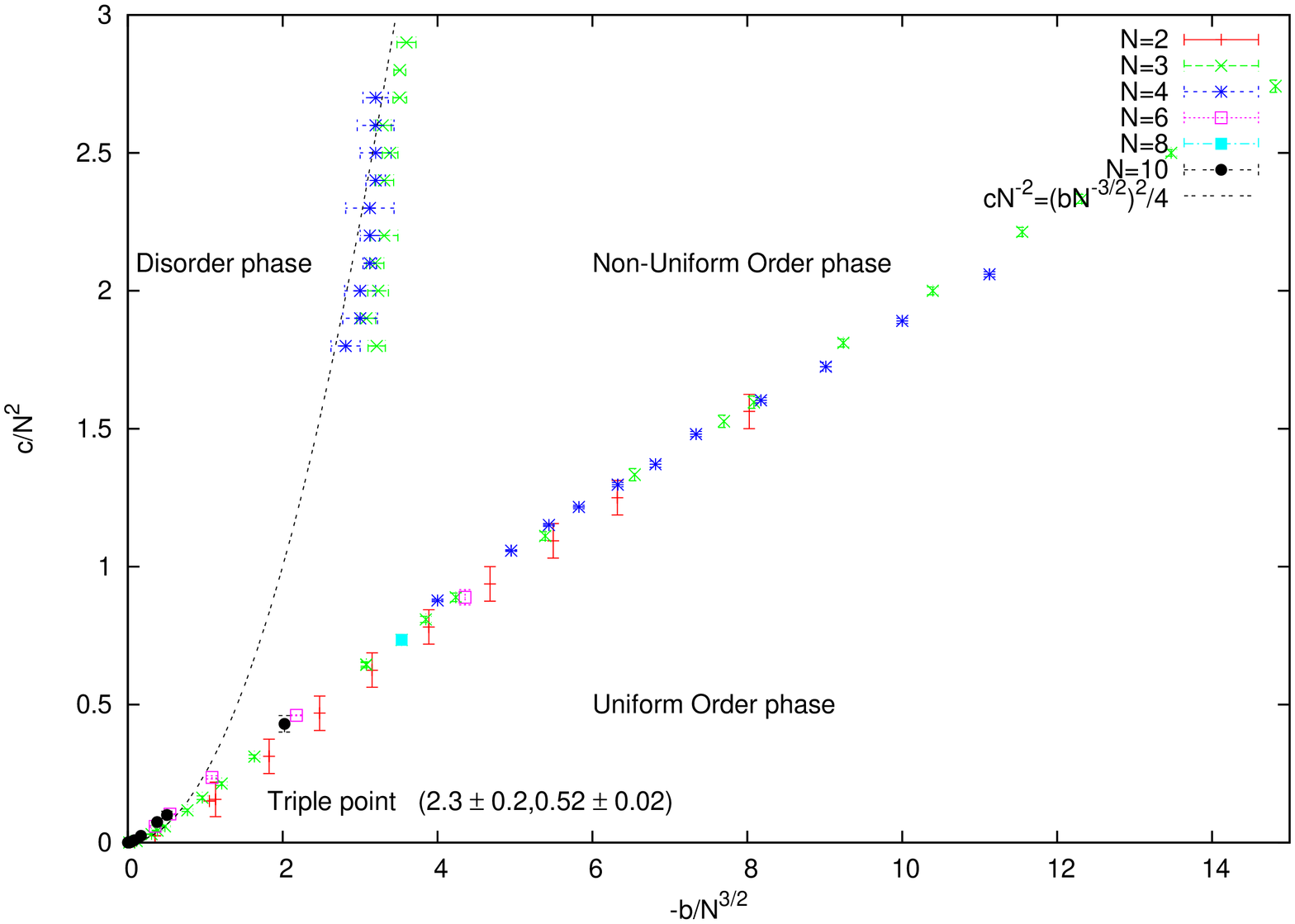}
\caption{The phase diagram of noncommutative phi-four theory on the fuzzy sphere. In the first figure the fits are reproduced from actual Monte Carlo data \cite{Ydri:2014rea}. Second figure reproduced from \cite{GarciaFlores:2009hf} with the gracious permission of  D.~O'Connor.}\label{phase_diagram}
\end{center}
\end{figure}

\begin{enumerate}
\item {\bf Model}~{\bf I}:~{\bf Model}~{\bf of}~{\bf \cite{O'Connor:2007ea}:}
\begin{itemize}
\item {\bf Ising:} Some of the results for the Ising transition for this model are shown on table (\ref{tablehis}). The critical point is taken at the peak of the susceptibility. The behavior of various observables is shown on figure (\ref{fig1}). The fit for the extrapolated critical value is given by
\begin{eqnarray}
\tilde{C}=0.291(0).(-\tilde{B})+0.104(1).\label{Ising}
\end{eqnarray}
This transition can be confirmed to be between disordered and uniform-ordered by looking at the eigenvalues distribution. In the disordered phase we have one-cut symmetric around zero whereas in the uniform-ordered we have one-cut symmetric around $\sqrt{-B/2C}$. See figure (\ref{fig12}).
\item{\bf Critical}~{\bf Exponents}: The calculation of the critical exponents of the above Ising transition is a very delicate exercise in Monte Carlo due to the known problem of critical slowing down and as a consequence the use of a different algorithm, for large values of $N$, such as the Wolf algorithm \cite{Wolff:1988uh} is essential. For values of $N$ less than $N=60$ the current algorithm is sufficient. In any case, this lengthy calculation is reported elsewhere. Suffice to say here that the critical exponents obtained are consistent, within the best statistical errors, with the Onsager solution of the Ising model in two dimensions given by the celebrated values \cite{Onsager:1943jn}

  \begin{eqnarray}
\nu=1~,~\beta=1/8~,~\gamma=7/4~,~\alpha=0~,~\eta=1/4.
\end{eqnarray} 
 This has always been known to be true but this is the first Monte Carlo direct calculation of these critical exponents.
\item {\bf Matrix:} Some of the results for the matrix transition between disorder and non-uniform order for this model are shown on table (\ref{tablehis2}). The critical point is determined at the point where the eigenvalue distributions go from one-cut in the disorder phase to  two-cut in the non-uniform phase. The splitting of the distribution is considered to have been occurred  when the hight of the distribution at $\lambda=0$ is less than some tolerance ${\rm Tol}$. We take ${\rm Tol}=0.001$. The behavior of the specific heat across this transition is effectively that of the pure quartic matrix model $a=0$. A sample of the corresponding specific heats and eigenvalue distributions is shown on figure (\ref{fig2}). The fit for the extrapolated critical value is given by
\begin{eqnarray}
\tilde{C}=2.206(67).(-\tilde{B})-7.039(301).\label{matrix}
\end{eqnarray}

\item {\bf Stripe:} This transition is quite difficult to observe in Monte Carlo even in this simplified setting which involves the sampling of $N$ eigenvalues. We can observe this transition for medium values of $\tilde{C}$ immediately above, but not too close to, the triple point.  The transition point is taken at the point where we observe a jump or a discontinuity in the zero power $P_0$ and the specific heat as seen on  figure (\ref{fig3}). 

Alternatively, we can approach the critical boundary by fixing the value of $\tilde{B}$ and changing $\tilde{C}$ starting from small values, i.e. inside the uniform ordered phase, until the curves for the total and  zero powers start to diverge marking the transition to the non-uniform ordered phase. The signal we obtain in this way is quite clear and unambiguous as shown on figure (\ref{fig3e}) and some measurements are included in table (\ref{tablehis3}).
Since this is a very delicate transition we do not perform any extrapolation of the critical point and the critical boundary is given by the fit of the largest value of $N$. In any case we observe no strong dependence on $N$ of the measured critical value $\tilde{C}$ as seen on table (\ref{tablehis3}).  The stripe critical line is then approximated by the fit for $N=50$ given by
\begin{eqnarray}
\tilde{C}=0.154(22).(-\tilde{B})+0.530(131)~,~N=50.\label{stripe}
\end{eqnarray}
\item {\bf Triple}~{\bf Point}~{\bf and}~{\bf Phase}~{\bf Diagram:} The location of the triple point is obtained from the intersection point of the Ising and matrix lines (\ref{Ising}) and (\ref{matrix}) respectively. Indeed,  the measurement of these two lines is more robust than the measurement of the stripe line (\ref{stripe}). We get then immediately 
 \begin{eqnarray}
(\tilde{B},\tilde{C})=(-3.73,1.19).
\end{eqnarray}
The phase diagram of the multitrace model of  \cite{O'Connor:2007ea} is shown on figure (\ref{pd}). The Ising and matrix transition data points are not shown explicitly but we only include their extrapolated fits whereas the $N=50$, $36$ and $25$ stripe data points are indicated explicitly.  We observe that the matrix boundary is closer to the doubletrace theory than it is to the quartic matrix model. The stripe critical boundary is of course expected to be closer to the $N=50$ measurement. 

\item {\bf Even}~{\bf Model}: This is the model in which we set all odd moments to zero in the action. We get then the doubletrace model
\begin{eqnarray}
V&=&BTr M^2+CTr M^4+D (Tr M^2)^2.
\end{eqnarray}
The most fundamental property of this model, observed in Monte Carlo simulation, is the absence of the uniform ordered phase. Indeed, only the disorder and the non-uniform order phases exist in the phase diagram. The critical boundary is very close to the doubletrace critical line shown on figure (\ref{pd}) which consists of two branches. Some precise measurements for the first branch are included in table  (\ref{tablehis4}). The turning point, towards the second branch, occurs around $\tilde{C}\sim 0.1$ where the critical point $-\tilde{B}_*$ becomes increasing, instead of decreasing, as we decrease $\tilde{C}$.
\end{itemize}
\item {\bf Model}~{\bf II}:~{\bf Model}~{\bf of}~{\bf \cite{Ydri:2014uaa}:} This model as pointed out previously is the correct approximation of noncommutative scalar $\Phi_2^4$ on the fuzzy sphere. However, this model is characterized by the absence of the uniform ordered phase and only the matrix transition line separating disordered and non-uniform ordered phases exists in the phase diagram. This fundamental result holds with and without odd terms. The role of the odd terms seems to be negligible and the two cases with and without odd terms are close. The doubletrace theory is also a very good approximation. A phase diagram is attached on figure  (\ref{pd}).

 The second fundamental observation in this case is the existence of a termination point. The matrix critical line does not extend to the origin and terminates at a point around $\tilde{C}=0.083$ in the case without odd terms, which agrees with the doubletrace theory prediction (\ref{pre}), and at a point around $\tilde{C}=0.4$ in the case with odd terms. This termination point is exhibited in Monte Carlo by the failure of the  Schwinger-Dyson identity (\ref{sd}). 

In particular, we observe for $N=50$ that  $\tilde{C}=0.4$ is the smallest value at which the disordered (one-cut) and non-uniform ordered (two-cut) phases are well defined. For $\tilde{C}=0.2-0.3$ the non-uniform ordered phase can not be clearly observed whereas for $\tilde{C}\leq 0.1$ both the disordered and the non-uniform ordered phases become indiscernible.  It is therefore natural to identify the triple point with the termination point  $\tilde{C}=0.4$. Our estimation of the termination point is given by
\begin{eqnarray}
(\tilde{B},\tilde{C})=(-1.05,0.4).
\end{eqnarray}
\end{enumerate}
\subsection{Grosse-Wulkenhaar Model}

The multitrace approach can also be applied to a regularized noncommutative $\Phi_2^4$ on the Moyal-Weyl plane in the matrix basis \cite{Ydri:2014uaa} with action given by
\begin{eqnarray}
S&=& Tr_N\bigg[\frac{1}{2}m^2M^2+\frac{u}{N}M^4+a\bigg(E M^2+\sqrt{\omega}\Gamma^+M\Gamma M\bigg)\bigg].
\end{eqnarray}
Two cases are of importance to us here:
\begin{enumerate}
\item The noncommutative theory without a harmonic oscillator term. In this case the effective action takes the form
\begin{eqnarray}
S_{\rm effe}&=&b Tr_N M^2+c Tr_N M^4+d (Tr_N M^2)^2+b_1 (Tr_N M)^2+c_1 (Tr_N M)^4\nonumber\\
&+&d_1 Tr_N M^2 (Tr_N M)^2+e Tr_N M Tr_N M^3.\nonumber\\
\end{eqnarray}
The parameters are given by
\begin{eqnarray}
&&b=\frac{m^2}{2}+\frac{aN}{2}~,~c=\frac{u}{N}-\frac{a^2N}{24}~,~d=-\frac{a^2}{12}\nonumber\\
&& b_1=-\frac{a}{2}~,~c_1=\frac{a^2}{24N^2}~,~d_1=-\frac{a^2}{12 N}~,~e=\frac{a^2}{6}.
\end{eqnarray}
If we assume the symmetry $M\longrightarrow -M$ then all odd moments vanish identically and we end up with the action 
\begin{eqnarray}
S_{\rm effe}=b Tr_N M^2+c Tr_N M^4+d (Tr_N M^2)^2.\label{gw1}
\end{eqnarray}
\item
At the self-dual point we have $\Omega^2=1$, and thus $\sqrt{\omega}=0$, and as a consequence the effective action reduces to the multitrace model
\begin{eqnarray}
S_{\rm effe}=b Tr_N M^2+c Tr_N M^4+d (Tr_N M^2)^2.\label{gw2}
\end{eqnarray}
The parameters $b$, $c$ and $d$ are given by
\begin{eqnarray}
b=\frac{m^2}{2}+\frac{aN}{2}~,~c=\frac{u}{N}-\frac{a^2N}{24}~,~d=\frac{a^2}{24}.
\end{eqnarray}
\end{enumerate}
Both the actions (\ref{gw1}) and (\ref{gw2}) do not contain odd moments and thus the corresponding phase diagrams are expected to not contain the uniform ordered phase with all matrix-like behavior as consequence.

 \begin{table}[H]
\begin{center}
\begin{tabular}{ |c|c|c|c|c|c| } 
 \hline
 $\tilde{C}$ & $N=10$ & $N=25$ & $N=36$ & $N=50$ & $\tilde{B}$  extrapolated\\ 
 \hline
 $0.3$ & -0.71  & -0.69 &-0.68 & -0.68 &  -0.672(2)   \\ 
 \hline
 $0.5$ & -1.44 & -1.39 &-1.38 & -1.38 &  -  1.361(3)    \\
 \hline
 $1.0$ &-3.21 & -3.13 &-3.11 & -3.10 &  -3.073(2)\\
 \hline
$1.2$ & -3.9 & -3.82 &-3.8 & -3.79 & -3.763(2)\\
 \hline
\end{tabular}
\end{center}
\caption{The Ising transition points for $N=10-50$. These are determined at the peak of the susceptibility (discontinuity in the specific heat). The search step is $0.01$.  }\label{tablehis}
\end{table}


\begin{table}[H]
\begin{center}
\begin{tabular}{ |c|c|c|c|c|c| } 
 \hline
 $\tilde{C}$ & $N=10$ & $N=25$ & $N=36$ & $N=50$ & $\tilde{B}$  extrapolated\\ 
 \hline
 $2.0$ & -4.925  & -4.475 &-4.325 & -4.225 & -4.090(37) \\ 
 \hline
 $2.5$ & -5.225 & -4.725 &-4.575 & -4.475 & -4.321(31)  \\
 \hline
 $3.0$ &-5.525 & -4.975 &-4.875 & -4.725 & -4.576(36)\\
 \hline
$4.0$ & -6.025 & -5.525 &-5.225 & -5.175 & -4.993(83) \\
 \hline
\end{tabular}
\end{center}
\caption{The matrix transition points for $N=10-50$. These are determined at the point where the eigenvalue distribution splits which is taken at the value  of $\tilde{B}$ where the distribution drops below 0.001 at zero. The search step is $0.025$. }\label{tablehis2}
\end{table}


\begin{table}[H]
\begin{center}
\begin{tabular}{ |c|c|c|c|c| } 
 \hline
 $\tilde{B}$ & $N=10$ & $N=25$ & $N=36$ & $N=50$ \\ 
 \hline
 $-9.0$ &$ 1.95\pm0.55$  &$ 1.8\pm0.4$ &$ 1.75\pm0.15$ &$ 1.95\pm0.15$ \\ 
 \hline
 $-8.0$ &$ 1.85\pm0.15$ &$ 1.65\pm0.25$ &$ 1.8\pm0.2$ &$ 1.8\pm0.2$  \\
 \hline
 $-7.0$ &$ 1.6\pm0.4$ &$ 1.45\pm0.05$ &$ 1.55\pm0.15$ &$ 1.6\pm0.3$ \\
 \hline
$-6.5$ &$ 1.65\pm0.15$ &$ 1.45\pm0.05$ & $ 1.45\pm0.05$ &$ 1.45 \pm0.15$\\
 \hline
$-6.0$ &$ 1.65\pm 0.15$ & $1.35\pm0.05$ &$1.55\pm0.05$ &$ 1.5\pm0.2$  \\
\hline
$-5.5$ &$ 1.55\pm 0.05$ & $1.35\pm0.05$ &$1.35\pm0.15$ &$ 1.35\pm0.05$  \\
\hline
$-5.0$ &$ 1.45\pm 0.05$ & $1.45\pm0.05$ &$1.25\pm0.15$ &$ 1.4\pm0.1$  \\
\hline
\end{tabular}
\end{center}
\caption{The non-uniform-to-uniform transition points $\tilde{C}$ for $N=10-50$. These are determined at the discontinuity of the zero power. }\label{tablehis3}
\end{table}


\begin{table}[H]
\begin{center}
\begin{tabular}{ |c|c|c|c|c|c|c| } 
 \hline
 $\tilde{C}$ & $N=10$ & $N=17$ & $N=25$ & $N=36$ &$N=50$ & $\tilde{B}$\\ 
 \hline
 $0.5$ & $-4.325$  & $-4.225$ &$-4.025$ & $-3.875$ &$- 3.825$ &  $-3.738(72)$\\ 
 \hline
 $1$ & $-4.375$ & $-4.175$ &$-4.025$ & $-3.875$ &$-3.775$ &  $-  3.685(54)$    \\
 \hline
 $2$ &$-4.925$ & $-4.675$ &$-4.475$ & $-4.325$ &$-4.225$ & $- 4.098(50)$  \\
 \hline
$3$ & $-5.475$ & $-5.275$ &$-5.025$ & $-4.825$ &$-4.725$& $-4.601(84)$ \\
 \hline
$5$ & $-6.525$ & $-6.275$ &$-6.025$ & $-5.725$ &$-5.625$&  $- 5.479(108)$  \\
\hline
\end{tabular}
\end{center}
\caption{The matrix transition points for $N=10-50$ in model I without odd terms. In this case only this transition exists and extends to a turning point in accordance with the doubletrace theory. The search step is $0.025$.}\label{tablehis4}
\end{table}


\section{Conclusion}
A Monte Carlo study of the multitrace quartic matrix model of \cite{O'Connor:2007ea} , which is claimed to be the first non-trivial correction to noncommutative $\Phi^4$ on the fuzzy sphere,  is presented. This model does not suffer from the severe ergodic problems encountered in the simulations of noncommutative $\Phi^4$ on the fuzzy sphere and the Metropolis algorithm is very effective in probing the entire phase space. In particular, Monte Carlo measurement of the one-cut-to-two-cut and the Ising transition lines as well as a direct Monte Carlo measurement of the non-uniform-to-uniform transition line are performed. The odd terms in the action which are dropped in \cite{O'Connor:2007ea} do play the central role in generating the Ising phase and the non-uniform-to-uniform transition line  and thus a triple point. A quantitative sketch of the phase diagram and the triple point is outline. 

The closely related multitrace quartic matrix model of \cite{Ydri:2014uaa},  which is  the correct approximation of noncommutative scalar $\Phi_2^4$ on the fuzzy sphere, is also considered in this article where it is shown that the one-cut-to-two-cut transition line does not extend to the origin and terminates at a point consistent with the triple point of noncommutative $\Phi^4$ on the fuzzy sphere \cite{GarciaFlores:2009hf}. 

We also commented in this article on the Grosse-Wulkenhaar model which is examined using a combination of the multitrace technique and the Monte Carlo method. At this order of the multitrace approximation the two models obtained in this case do not exhibit  the Ising phase and the non-uniform-to-uniform transition line.

\appendix
\section{Susceptibility and Specific Heat} 
\subsection{Susceptibility}
We consider $\Phi^4$ on the fuzzy sphere coupled to a constant magnetic field $H$ given by the action
\begin{eqnarray}
S=Tr \big(a \Phi[L_a,[L_a,\Phi]]+b\Phi^2+c\Phi^4+H \Phi\big).
\end{eqnarray}
The magnetization and the susceptibility are defined by
\begin{eqnarray}
{\rm magnetization}&=&\frac{1}{N}<Tr\Phi>\nonumber\\
&=&-\frac{1}{N}\frac{\partial}{\partial H}\ln Z.
\end{eqnarray}

\begin{eqnarray}
{\rm susceptibility}&=&<(Tr\Phi)^2>-<Tr\Phi>^2\nonumber\\
&=&\frac{\partial^2}{\partial H^2}\ln Z\nonumber\\
&=&-N\frac{\partial}{\partial H}{\rm magnetization}.
\end{eqnarray}
On the fuzzy sphere we have
\begin{eqnarray}
x_a=\frac{2R}{N}L_a~,~[x_a,x_b]=\frac{i\theta}{R}\epsilon_{abc}x_c~,~\theta=\frac{2R^2}{N}~,~Tr=\frac{N}{4\pi R^2}\int d^2x.
\end{eqnarray}
The regularized noncommutative plane is then defined by 
\begin{eqnarray}
x_3=R~,~[x_1,x_2]=i\theta~,~\partial_i=-\frac{1}{R}\epsilon_{ij}L_j=-\frac{1}{\theta}\epsilon_{ij}x_j~,~\int d^2x=2\pi\theta Tr.
\end{eqnarray}
We have $\epsilon_{12}=1$. The above action becomes, including a rescaling of the field $\Phi\longrightarrow \phi=\sqrt{Na/2\pi}\Phi$, given by the equation
\begin{eqnarray}
S=2\pi\theta Tr \big(\frac{1}{2} \phi\partial_i\partial_i\phi+\frac{1}{2}m^2\phi^2+\frac{1}{4}\lambda\phi^4+h \phi\big).
\end{eqnarray}
\begin{eqnarray}
m^2=\frac{b}{aR^2}~,~\lambda=\frac{4\pi c}{Na^2R^2}~,~h=\sqrt{\frac{N}{2\pi a}}\frac{H}{2R^2}.
\end{eqnarray}
The commutative limit is $\theta\longrightarrow 0$. By using a lattice in this limit we have
 \begin{eqnarray}
S=l^2\sum_n \big(\frac{1}{2} (\phi\partial_i\partial_i\phi)_{\rm lattice}+\frac{1}{2}m^2\phi_n^2+\frac{1}{4}\lambda\phi_n^4+h \phi_n\big).
\end{eqnarray}
We compute in this limit on the lattice
\begin{eqnarray}
{\rm magnetization}&=&\frac{1}{N}<Tr\Phi>\nonumber\\
&\longrightarrow& \frac{{\cal N}^2l^2}{4\pi R^2}<\frac{1}{{\cal N}^2}\sum_n\Phi_n>.
\end{eqnarray}
The volume of the lattice must be equal to the area of the sphere, viz  ${\cal N}^2l^2=4\pi R^2$. Also we compute
\begin{eqnarray}
{\rm susceptibility}&=&<(Tr\Phi)^2>-<Tr\Phi>^2\nonumber\\
&=&\frac{N^2}{{\cal N}^4}\frac{{\cal N}^2l^2}{4\pi R^2}\big(<(\sum_n\Phi_n)^2>-<\sum_n\Phi_n>^2\big).
\end{eqnarray}
\subsection{Specific Heat}
The specific heat is defined by
\begin{eqnarray}
C_v&=&\frac{\partial^2}{\partial\beta^2}\ln Z\nonumber\\
&=&<S^2>-<S>^2.
\end{eqnarray}
The inverse temperature is introduced in the usual way as
 \begin{eqnarray}
Z=\int dM \exp(-\beta S[M]).
\end{eqnarray}
The calculation of the effective potential proceeds as before with the replacement $a\longrightarrow a\beta$. The partition function in the quartic multitrace approximation is

\begin{eqnarray}
Z
&=&\int d \Lambda~\Delta^2(\Lambda) ~\exp\big(-\beta V_0 \big)+\beta \int d \Lambda~\Delta^2(\Lambda) ~\exp\big(-\beta V_0 \big)\big(-V_2\big)\nonumber\\
&+&\beta^2\int d \Lambda~\Delta^2(\Lambda) ~\exp\big(-\beta V_0 \big)\big(-V_4+\frac{1}{2}V_2^2 \big).
\end{eqnarray}
A straightforward calculation yields 
\begin{eqnarray}
C_v
&=&<(V+V_4)^2>-<(V+V_4)>^2-2<\big(V_4+2V_4^2+2V_2V_4\big)>.
\end{eqnarray}
The last term could make this approximation of the specific heat negative. This actually happens in the approximation of  \cite{O'Connor:2007ea}.

\paragraph{Acknowledgment:}  
This research was supported by CNEPRU: "The National (Algerian) Commission for the Evaluation of
University Research Projects"  under contract number ${\rm DO} 11 20 13 00 09$.


\begin{thebibliography}{10}

\bibitem{Gubser:2000cd} 
  S.~S.~Gubser and S.~L.~Sondhi,
  ``Phase structure of noncommutative scalar field theories,''
  Nucl.\ Phys.\ B {\bf 605}, 395 (2001)
  [hep-th/0006119].

\bibitem{Ambjorn:2002nj} 
  J.~Ambjorn and S.~Catterall,
  ``Stripes from (noncommutative) stars,''
  Phys.\ Lett.\ B {\bf 549}, 253 (2002)
  [hep-lat/0209106].

\bibitem{Ambjorn:2000cs}
  J.~Ambjorn, Y.~M.~Makeenko, J.~Nishimura and R.~J.~Szabo,
  ``Lattice gauge fields and discrete noncommutative Yang-Mills theory,''
  JHEP {\bf 0005}, 023 (2000)
  [arXiv:hep-th/0004147].

\bibitem{Grosse:2003nw}
  H.~Grosse and R.~Wulkenhaar,
  ``Renormalisation of phi**4 theory on noncommutative R**2 in the matrix
  base,''
  JHEP {\bf 0312}, 019 (2003)
  [arXiv:hep-th/0307017].

\bibitem{Hoppe:1982}
  J.~Hoppe,
  ``Quantum theory of a massless relativistic surface and a two-dimensional bound state problem,''
  Ph.D thesis,MIT,1982. 


\bibitem{Madore:1991bw}
  J.~Madore,
  ``The fuzzy sphere,''
  Class.\ Quant.\ Grav.\  {\bf 9}, 69 (1992).

\bibitem{Connes:1994yd}
  A.~Connes,
  ``Noncommutative geometry,''
   Academic Press,London, 1994.

\bibitem{Frohlich:1993es}
  J.~Frohlich and K.~Gawedzki,
  ``Conformal field theory and geometry of strings,''
  arXiv:hep-th/9310187.

\bibitem{GarciaFlores:2009hf} 
  F.~Garcia Flores, X.~Martin and D.~O'Connor,
  ``Simulation of a scalar field on a fuzzy sphere,''
  Int.\ J.\ Mod.\ Phys.\ A {\bf 24}, 3917 (2009)
  [arXiv:0903.1986 [hep-lat]]. 

\bibitem{GarciaFlores:2005xc} 
  F.~Garcia Flores, D.~O'Connor and X.~Martin,
  ``Simulating the scalar field on the fuzzy sphere,''
  PoS LAT {\bf 2005}, 262 (2006)
  [hep-lat/0601012].

\bibitem{Martin:2004un} 
  X.~Martin,
  ``A Matrix phase for the phi**4 scalar field on the fuzzy sphere,''
  JHEP {\bf 0404}, 077 (2004)
  [hep-th/0402230].

\bibitem{Panero:2006bx} 
  M.~Panero,
  ``Numerical simulations of a non-commutative theory: The Scalar model on the fuzzy sphere,''
  JHEP {\bf 0705}, 082 (2007)
  [hep-th/0608202].

\bibitem{Medina:2007nv} 
  J.~Medina, W.~Bietenholz and D.~O'Connor,
  ``Probing the fuzzy sphere regularisation in simulations of the 3d lambda phi**4 model,''
  JHEP {\bf 0804}, 041 (2008)
  [arXiv:0712.3366 [hep-th]].


\bibitem{Das:2007gm} 
  C.~R.~Das, S.~Digal and T.~R.~Govindarajan,
  ``Finite temperature phase transition of a single scalar field on a fuzzy sphere,''
  Mod.\ Phys.\ Lett.\ A {\bf 23}, 1781 (2008)
  [arXiv:0706.0695 [hep-th]].



\bibitem{O'Connor:2007ea} 
  D.~O'Connor and C.~Saemann,
  ``Fuzzy Scalar Field Theory as a Multitrace Matrix Model,''
  JHEP {\bf 0708}, 066 (2007)
  [arXiv:0706.2493 [hep-th]].

\bibitem{Saemann:2010bw} 
  C.~Saemann,
  ``The Multitrace Matrix Model of Scalar Field Theory on Fuzzy CP**n,''
  SIGMA {\bf 6}, 050 (2010)
  [arXiv:1003.4683 [hep-th]].



\bibitem{Polychronakos:2013nca} 
  A.~P.~Polychronakos,
  ``Effective action and phase transitions of scalar field on the fuzzy sphere,''
  arXiv:1306.6645 [hep-th].

\bibitem{Tekel:2014bta} 
  J.~Tekel,
  ``Uniform order phase and phase diagram of scalar field theory on fuzzy CP**n,''
  arXiv:1407.4061 [hep-th].

\bibitem{Nair:2011ux} 
  V.~P.~Nair, A.~P.~Polychronakos and J.~Tekel,
  ``Fuzzy spaces and new random matrix ensembles,''
  Phys.\ Rev.\ D {\bf 85}, 045021 (2012)
  [arXiv:1109.3349 [hep-th]].

\bibitem{Tekel:2013vz} 
  J.~Tekel,
  ``Random matrix approach to scalar fields on fuzzy spaces,''
  Phys.\ Rev.\ D {\bf 87}, no. 8, 085015 (2013)
  [arXiv:1301.2154 [hep-th]].


\bibitem{Bietenholz:2004xs} 
  W.~Bietenholz, F.~Hofheinz and J.~Nishimura,
  ``Phase diagram and dispersion relation of the noncommutative lambda phi**4 model in d = 3,''
  JHEP {\bf 0406}, 042 (2004)
  [hep-th/0404020].

\bibitem{Lizzi:2012xy} 
  F.~Lizzi and B.~Spisso,
  ``Noncommutative Field Theory: Numerical Analysis with the Fuzzy Disc,''
  Int.\ J.\ Mod.\ Phys.\ A {\bf 27}, 1250137 (2012)
  [arXiv:1207.4998 [hep-th]].


\bibitem{Mejia-Diaz:2014lza} 
  H.~Mejía-Díaz, W.~Bietenholz and M.~Panero,
  ``The Continuum Phase Diagram of the 2d Non-Commutative lambda phi**4 Model,''
  arXiv:1403.3318 [hep-lat].






\bibitem{Ydri:2014uaa} 
  B.~Ydri,
  ``A Multitrace Approach to Noncommutative $\Phi_2^4$,''
  arXiv:1410.4881 [hep-th].



\bibitem{Ydri:2014rea} 
  B.~Ydri,
  ``New algorithm and phase diagram of noncommutative $\phi^4$ on the fuzzy sphere,''
  JHEP {\bf 1403}, 065 (2014)
  [arXiv:1401.1529 [hep-th]].


\bibitem{Minwalla:1999px}
  S.~Minwalla, M.~Van Raamsdonk and N.~Seiberg,
  ``Noncommutative perturbative dynamics,''
  JHEP {\bf 0002}, 020 (2000)
  [arXiv:hep-th/9912072].

\bibitem{brazovkii}
  S.~A.~Brazovkii,
  ``Phase Transition of an Isotropic System to a Nonuniform State,''
  Zh. Eksp. Teor. Fiz {\bf 68}, (1975) 175-185.









\bibitem{Shimamune:1981qf} 
  Y.~Shimamune,
  ``On The Phase Structure Of Large N Matrix Models And Gauge Models,''
  Phys.\ Lett.\ B {\bf 108}, 407 (1982).

\bibitem{eynard}
  B.~Eynard,
  ``Random Matrices,''
  Cours de Physique Theorique de Saclay.

\bibitem{Brezin:1977sv} 
  E.~Brezin, C.~Itzykson, G.~Parisi and J.~B.~Zuber,
  ``Planar Diagrams,''
  Commun.\ Math.\ Phys.\  {\bf 59}, 35 (1978).

\bibitem{Kawahara:2007eu} 
  N.~Kawahara, J.~Nishimura and A.~Yamaguchi,
  ``Monte Carlo approach to nonperturbative strings - Demonstration in noncritical string theory,''
  JHEP {\bf 0706}, 076 (2007)
  [hep-th/0703209].

\bibitem{Steinacker:2005wj} 
  H.~Steinacker,
  ``A Non-perturbative approach to non-commutative scalar field theory,''
  JHEP {\bf 0503}, 075 (2005)
  [hep-th/0501174].

\bibitem{Onsager:1943jn} 
  L.~Onsager,
  ``Crystal statistics. 1. A Two-dimensional model with an order disorder transition,''
  Phys.\ Rev.\  {\bf 65}, 117 (1944).


 
\bibitem{Wolff:1988uh} 
  U.~Wolff,
  ``Collective Monte Carlo Updating for Spin Systems,''
  Phys.\ Rev.\ Lett.\  {\bf 62}, 361 (1989).

\bibitem{DelgadilloBlando:2007vx} 
  R.~Delgadillo-Blando, D.~O'Connor and B.~Ydri,
  ``Geometry in Transition: A Model of Emergent Geometry,''
  Phys.\ Rev.\ Lett.\  {\bf 100}, 201601 (2008)
  [arXiv:0712.3011 [hep-th]].

\end{thebibliography}
\end{document}